\documentclass[altaffilletter,aps,nofootinbib,twocolumn,prd,eqsecnum,preprintnumbers,superscriptaddress]{revtex4}
\pdfoutput=1
\usepackage[caption=false]{subfig}
\usepackage{graphicx}
\usepackage{amsmath}
\usepackage{amsfonts}
\usepackage{amssymb}
\usepackage{color}
\usepackage{bm}
\usepackage{mathrsfs}
\usepackage{epstopdf}
\usepackage{url}
\usepackage{footnote}
\usepackage{textcomp}
\usepackage{dsfont}
\usepackage{ulem}
\usepackage{hyperref}
\usepackage{enumerate}

\makeatletter
\newcommand*{\rom}[1]{\expandafter\@slowromancap\romannumeral #1@}
\makeatother

\begin{document}

\title{Gravitational perturbations of non-singular black holes in conformal gravity}

\author{Che-Yu Chen}
\email{b97202056@gmail.com}
\affiliation{Department of Physics and Center for Theoretical Sciences, National Taiwan University, Taipei, Taiwan 10617}
\affiliation{LeCosPA, National Taiwan University, Taipei, Taiwan 10617}

\author{Pisin Chen}
\email{pisinchen@phys.ntu.edu.tw}
\affiliation{Department of Physics and Center for Theoretical Sciences, National Taiwan University, Taipei, Taiwan 10617}
\affiliation{LeCosPA, National Taiwan University, Taipei, Taiwan 10617}
\affiliation{Kavli Institute for Particle Astrophysics and Cosmology, SLAC National Accelerator Laboratory, Stanford University, Stanford, CA 94305, USA}
\begin{abstract}
It is believed that in the near future, gravitational wave detections will become a promising tool not only to test gravity theories, but also to probe extremely curved spacetime regions in our universe, such as the surroundings of black holes. In this paper, we investigate the quasinormal modes (QNMs) of the axial gravitational perturbations of a class of non-singular black holes conformally related to the Schwarzschild black hole. These non-singular black holes can be regarded as the vacuum solution of a family of conformal gravity theories which are invariant under conformal transformations. After conformal symmetry is broken, these black holes produce observational signatures different from those of the Schwarzschild black hole, such as their QNM frequencies. We assume that the spacetime is described by the Einstein equation with the effective energy momentum tensor of an anisotropic fluid. The master equation describing the QNMs is derived, and the QNM frequencies are evaluated with the Wentzel-Kramers-Brillouin (WKB) method up to the 6th order. As expected, the QNM spectra of these non-singular black holes deviate from those of the Schwarzschild black hole, indicating the possibility of testing these black hole solutions with the help of future gravitational wave detections.
\end{abstract}

\maketitle

\section{Introduction}   
The direct detection of gravitational waves (GWs) from the coalescence of binary black holes \cite{Abbott:2016blz,Abbott:2017oio} is a milestone for the development of modern physics and astronomy. This achievement not merely validates Einstein's general relativity (GR) once again, but also ushers in a new era of GW detections worldwide. Furthermore, the GW signals emitted from the merger of binary neutron stars and the accompanied electromagnetic signals were detected recently, with an accurate localization of the source \cite{TheLIGOScientific:2017qsa}. Undoubtedly, we are currently ushered in a new era of GW and multi-messenger astronomy.

One of the important uses of GW detections is to test extended theories of gravity, or more particularly, to distinguish black hole solutions in different theories from their GR counterparts. Even thought GR has been serving as the best description of our universe so far, it still suffers from several unexplained puzzles. One of the examples is the prediction of spacetime singularities such as those inside the black holes. Intuitively, one would expect that near the singularity, the bending of the spacetime is too strong and the energy density is too huge, such that a quantum description of gravity is necessary as GR turns out to be inaccurate there (see Refs.~\cite{Ashtekar:2005qt,Nicolini:2005vd,LopezDominguez:2006wd,Ashtekar:2018lag,Ashtekar:2018cay,Hossenfelder:2009fc,Bojowald:2018xxu,BenAchour:2018khr,Bodendorfer:2019cyv} for some attempts of including quantum corrections to ameliorate black hole singularities). Since we still do not have a self-consistent and complete quantum theory of gravity, one can consider modifying (or extending) GR at high curvature regimes from a phenomenological point of view \cite{Capozziello:2011et}. Theoretically, there can be a huge class of extended theories of gravity and it is hoped that GW detections can shed light on testing, or even falsifying, them in the future.

In this paper, we will focus on another strategy devoting to the resolution of singularities. The idea is to make use of the conformal symmetry of the spacetime \cite{Englert:1976ep,narlikar:1977nf}. It can be expected that at high curvature regimes, the spacetime could preserve the conformal symmetry in the sense that it can be described by an underlying gravity theory which is invariant under following conformal transformations
\begin{equation}
\hat{g}_{\mu\nu}\rightarrow\hat{g}_{\mu\nu}=S(x)g_{\mu\nu}\,,
\end{equation}
where $S(x)$ is a conformal factor depending on the spacetime coordinates. There are several conformal gravity theories in the literature, for example, those constructed with an auxiliary scalar field \cite{tHooft:2011aa,Dabrowski:2008kx}, or with the Weyl curvature tensor \cite{Mannheim:2011ds,Mannheim:2016lnx}. In the symmetry phase where the theory is conformally invariant, the spacetime singularity can be easily removed after a suitable conformal transformation. In this regard, the spacetime singularity turns out to be an artifact of different choices of conformal gauges, just as the notion of coordinate singularities in GR, which can be removed with proper coordinate transformations. Following this direction, the authors of Refs.~\cite{Modesto:2016max,Bambi:2016wdn} proposed a class of non-singular black hole solutions which are conformally related either to the Schwarzschild black hole or to the Kerr black hole. The spacetime is geodescially complete and the curvature is finite everywhere for these solutions \cite{Bambi:2016wdn}. In this paper, we will follow a strategy similar to that in Ref.~\cite{Bambi:2016wdn}. Instead of considering a specific conformal gravity theory, we will treat the non-singular black holes proposed in \cite{Modesto:2016max,Bambi:2016wdn} as solutions to a family of conformal gravity theories. The results obtained here can be expected as a generic feature among this family of theories.

As mentioned in \cite{Bambi:2016wdn}, the current spacetime is obviously not conformally invariant. Therefore, if we believe that the conformal symmetry does play a crucial role in Nature and resolve the spacetime singularities, there must be a phase transition where the conformal symmetry is broken. At the phase transition, a particular preferred spacetime should be chosen from a huge number of conformally invariant spacetimes. After the symmetry breaking, different conformally related spacetimes result in different observational signatures. In Refs~\cite{Bambi:2017yoz,Zhou:2018bxk}, this class of non-singular rotating black holes has been tested by using X-ray observational data. The scalar field perturbations of the non-singular and non-rotating black holes were discussed in Ref.~\cite{Toshmatov:2017bpx}. The formation and evaporation of both the neutral and charged black holes have been studied in Refs.~\cite{Bambi:2016yne,Bambi:2017ott}. The properties of slowly rotating magnetized compact starts \cite{Turimov:2018pey}, and other non-singular spacetime metrics \cite{Chakrabarty:2017ysw} were investigated. In Ref.~\cite{Toshmatov:2017kmw}, the authors studied the violation of energy conditions for these black hole solutions and assumed that Nature would select the solutions which violate less amount of energy conditions at the symmetry breaking. A general study of rotating black holes in conformal gravity has been done in Ref.~\cite{Zhang:2018qdk}. See also Ref.~\cite{,Zhang:2017amt} for some interesting dynamical spacetimes in a particular conformal gravity theory. 

In order to test these non-singular black hole solutions, in this paper we will study the quasinormal modes (QNMs) of their axial gravitational perturbations. The GWs emitted at the final stage of a merger event, that is, the ringdown signals, are characterized by the QNMs and can be described by the theory of black hole perturbations. In this stage, the distorted black hole can be regarded as a dissipative system. The system has a discrete spectrum and the QNM frequencies are complex numbers, whose real part and imaginary part describe the oscillations of the perturbations and the decay of the amplitude, respectively. Since we do not consider a specific conformal gravity theory, we will derive the master equation of the perturbations assuming that the solution is governed by the Einstein equation coupled with the effective energy momentum tensor of an anisotropic fluid. We will exhibit that the QNM frequencies depend on the conformal factors. To calculate the QNM frequencies, the WKB method up to the 6th order is used \cite{Schutz:1985zz,Iyer:1986np,Konoplya:2003ii,Matyjasek:2017psv}. We would like to stress that testing gravity theories and black hole solutions by using QNMs has been a popular research direction recently, such as in the Horndeski gravity \cite{Kobayashi:2012kh,Kobayashi:2014wsa,Minamitsuji:2014hha,Dong:2017toi,Tattersall:2018nve}, metric $f(R)$ gravity \cite{Sebastian:2014qra,Bhattacharyya:2017tyc,Bhattacharyya:2018qbe}, Palatini type gravity \cite{Chen:2018mkf,Chen:2018vuw}, massive gravity \cite{Fernando:2014gda,Prasia:2016fcc}, Einstein-dilaton-Gauss-Bonnet gravity \cite{Blazquez-Salcedo:2016enn,Blazquez-Salcedo:2017txk,Blazquez-Salcedo:2018pxo,Blazquez-Salcedo:2016yka}, the Randall-Sundrum braneworld model \cite{Toshmatov:2016bsb}, Ho\v{r}ava-Lifshitz gravity \cite{Lin:2016wci}, higher dimensional black holes \cite{Cardoso:2003qd,Cardoso:2004cj,Cardoso:2003vt}, and Einstein-aether theory \cite{Ding:2017gfw}, etc. See Refs.~\cite{Nollert:1999ji,Berti:2009kk,Konoplya:2011qq,Berti:2015itd,Cardoso:2019mqo} for nice reviews on the latest progress of the field. 

This paper is outlined as follows. In section \ref{sec.nonbh}, we briefly review the conformal transformation introduced in Ref.~\cite{Bambi:2016wdn} in which the rescaled metric turns out to be a non-singular black hole solution. In section \ref{sec.axial}, we assume that the non-singular black hole is described by the Einstein equation with an effective energy momentum tensor, and present the master equation of the axial perturbations of the black hole. In section~\ref{sect.wkb}, we use the WKB method up to the 6th order to calculate the fundamental QNM frequencies. In this section, we also analyze the eikonal QNMs, and the asymptotic behaviors of the QNM frequencies when the parameters of the conformal factor are large. The time domain profiles of the perturbations and the late-time tails are discussed. We finally conclude in section \ref{conclu}. An appendix is included in an attempt to present the derivation of the master equation of the axial perturbations.

\section{Non-singular black holes in conformal gravity}\label{sec.nonbh}
In the family of conformal theories of gravity where the spacetime respects the conformal symmetry, all the spacetime metrics related through conformal transformations are physically equivalent. In this regard, the spacetime singularity can be removed after a suitable conformal transformation. The spacetime singularity turns out to be a mathematical artifact of choosing different conformal factors. In this section, we will briefly review the non-singular black hole metric proposed in Ref.~\cite{Bambi:2016wdn}.

In the conformal symmetry phase, the non-singular black hole metric is conformally related to the Schwarzschild black hole as follows \cite{Bambi:2016wdn}:
\begin{align}
ds^2&=S(r)ds^2_{\textrm{Schw}}\nonumber\\
&=-S(r)f(r)dt^2+\frac{S(r)dr^2}{f(r)}+S(r)r^2d\Omega^2\,,\label{nonsingmetric}
\end{align}
where $ds^2_{\textrm{Schw}}$ is the Schwarzschild line element. The metric function $f(r)$ is
\begin{equation}
f(r)=1-\frac{r_s}{r}\,,
\end{equation}
where $r_s$ is the Schwarzschild radius. In fact, there are many choices of $S(r)$ such that the spacetime described by
\begin{equation}
ds^2=\hat{g}_{\mu\nu}dx^\mu dx^\nu\,,
\end{equation}
is everywhere non-singular. We consider the following conformal factor which was introduced in \cite{Bambi:2016wdn}:
\begin{equation}
S(r)=\left(1+\frac{L^2}{r^2}\right)^{2N}\,,\label{confS}
\end{equation}
where $N$ is an arbitrary positive integer and $L$ is a new length scale. In the rest of this paper, we will use the following dimensionless rescalings:
\begin{equation}
\frac{L}{r_s}\rightarrow L\,,\qquad \frac{r}{r_s}\rightarrow r\,,
\end{equation} 
for the sake of convenience.

First of all, it can be seen that the conformal factor $S(r)$ reduces to unity when $L\ll r$. Therefore, in this limit the Schwarzschild solution is recovered. Interestingly, as long as $N\neq0$ and $L\neq0$, the spacetime is everywhere non-singular. This can be smattered by calculating curvature invariants of the spacetime. Actually, it can be shown that the curvature invariants are finite everywhere, including the origin $r=0$. For instance, the Ricci scalar and the Kretschmann scalar of the metric \eqref{nonsingmetric} near $r\rightarrow0$ can be approximated as follows \cite{Bambi:2016wdn}
\begin{equation}
R[\hat{g}]\equiv\hat{g}^{\mu\nu}R_{\mu\nu}[\hat{g}]\approx\frac{24N^2}{L^{4N}}r^{4N-3}\,,
\end{equation}
and
\begin{align}
\mathcal{K}[\hat{g}]&\equiv R_{\alpha\beta\gamma\delta}[\hat{g}]R^{\alpha\beta\gamma\delta}[\hat{g}]\nonumber\\&\approx\frac{12\left(1+12N^2-16N^3+16N^4\right)}{L^{8N}}r^{8N-6}\,,
\end{align}
respectively. Because $N$ is a positive integer, the curvature invariants are finite everywhere, including the origin. In Ref.~\cite{Bambi:2016wdn}, it was also shown that the spacetime described by the metric \eqref{nonsingmetric} is geodesically complete, justifying the non-singular property of the spacetime.

\section{Axial perturbations}\label{sec.axial}
As mentioned previously, the spacetime \eqref{nonsingmetric} is physically equivalent to the Schwarzschild metric in the conformal symmetry phase because they are different just by a conformal rescaling. However, the universe is currently not conformally invariant. There must be a phase transition where the conformal symmetric of the spacetime is broken and Einstein GR is then recovered. Theoretically, There should be a particular selection mechanism at the symmetry breaking to pick a spacetime metric from those infinite metrics which are conformally invariant at the symmetry phase. If it is the non-singular black hole metric \eqref{nonsingmetric} that has been chosen at the symmetry breaking, instead of the Schwarzschild one, we can then distinguish them observationally. In this paper, we will investigate the QNMs of the axial perturbations of the non-singular black hole \eqref{nonsingmetric}. 

The QNMs generated by a massless scalar field and those by electromagnetic perturbations have been studied in Ref.~\cite{Toshmatov:2017bpx}. In that paper, the authors neglected the back reaction of the fields on the spacetime. Furthermore, in the absence of a specific underlying theory, the authors derived the master equations by assuming the validity of the Klein-Gordon equation and the Maxwell equation. It should be noticed that the conservation of these additional test fields may not be satisfied in some gravitational theories, unless the matter sector can be guaranteed to be minimally coupled with the metric $\hat{g}_{\mu\nu}$. 

In order to study the axial gravitational perturbations, one needs to perturb the gravitational equations as well as the energy momentum tensor. Again, in the absence of the underlying theory, an alternative method should be applied. In this paper, we will assume that the solution is governed by the Einstein equation with an effective energy momentum tensor. In Ref.~\cite{Toshmatov:2017kmw}, the authors have used this approach to address the energy conditions of the non-singular black holes in the conformal gravity. Similar method has also been applied to study black hole solutions with quantum corrections in several literature \cite{Nicolini:2005vd,Ashtekar:2018lag,Ashtekar:2018cay}. From a phenomenological point of view, the effective energy momentum tensor corresponding to the black hole solution \eqref{nonsingmetric} can be described by an anisotropic fluid:
\begin{equation}
T_{\mu\nu}=\left(\rho+p_2\right)u_\mu u_\nu+\left(p_1-p_2\right)x_\mu x_\nu+p_2\hat{g}_{\mu\nu}\,,\label{anisopf}
\end{equation}
where $\rho$ is the energy density measured by a comoving observer with the fluid, and $u^\mu$ and $x^\mu$ are the timelike four-velocity and the spacelike unit vector orthogonal to $u^\mu$ and angular directions, respectively. On the expression \eqref{anisopf}, $p_1$ and $p_2$ are the radial pressure and the tangential pressure, respectively. Note that $u^\mu$ and $x^\mu$ satisfy
\begin{equation}
u_\mu u^\mu=-1\,,\qquad x_\mu x^\mu=1\,,\label{fourvelocity}
\end{equation}
where the indices are raised and lowered by the metric $\hat{g}_{\mu\nu}$. In the comoving frame, we can assume $u^\mu=\left(u^t,0,0,0\right)$ and $x^\mu=\left(0,x^r,0,0\right)$. From Eq.~\eqref{fourvelocity}, we have
\begin{equation}
u_t^2=\hat{g}_{tt}u_tu^t=-\hat{g}_{tt}\,,\qquad x_r^2=\hat{g}_{rr}x_rx^r=\hat{g}_{rr}\,,
\end{equation}
at the background level. The components of energy momentum tensor read
\begin{align}
T_{tt}&=-\hat{g}_{tt}\rho\,,\qquad T_t^t=-\rho\,,\\
T_{rr}&=\hat{g}_{rr}p_1\,,\qquad T_r^r=p_1\,,\\
&T_{\theta}^{\theta}=T_\phi^\phi=p_2\,.
\end{align}
The explicit expressions of $\rho$, $p_1$, and $p_2$ are functions of $r$ and they can be derived by calculating the corresponding Einstein tensor $G_{\mu\nu}(\hat{g})$ constructed from the metric \eqref{nonsingmetric}. In the appendix~\ref{deapp}, we will use the tetrad formalism \cite{Chandrabook} to derive the master equation of the axial perturbations of the metric \eqref{nonsingmetric}. It will be shown explicitly there that the exact forms of $\rho$, $p_1$, and $p_2$ have nothing to do with the master equation.

\begin{figure*}[tt]
\centering
\graphicspath{{fig/}}
\includegraphics[scale=0.46]{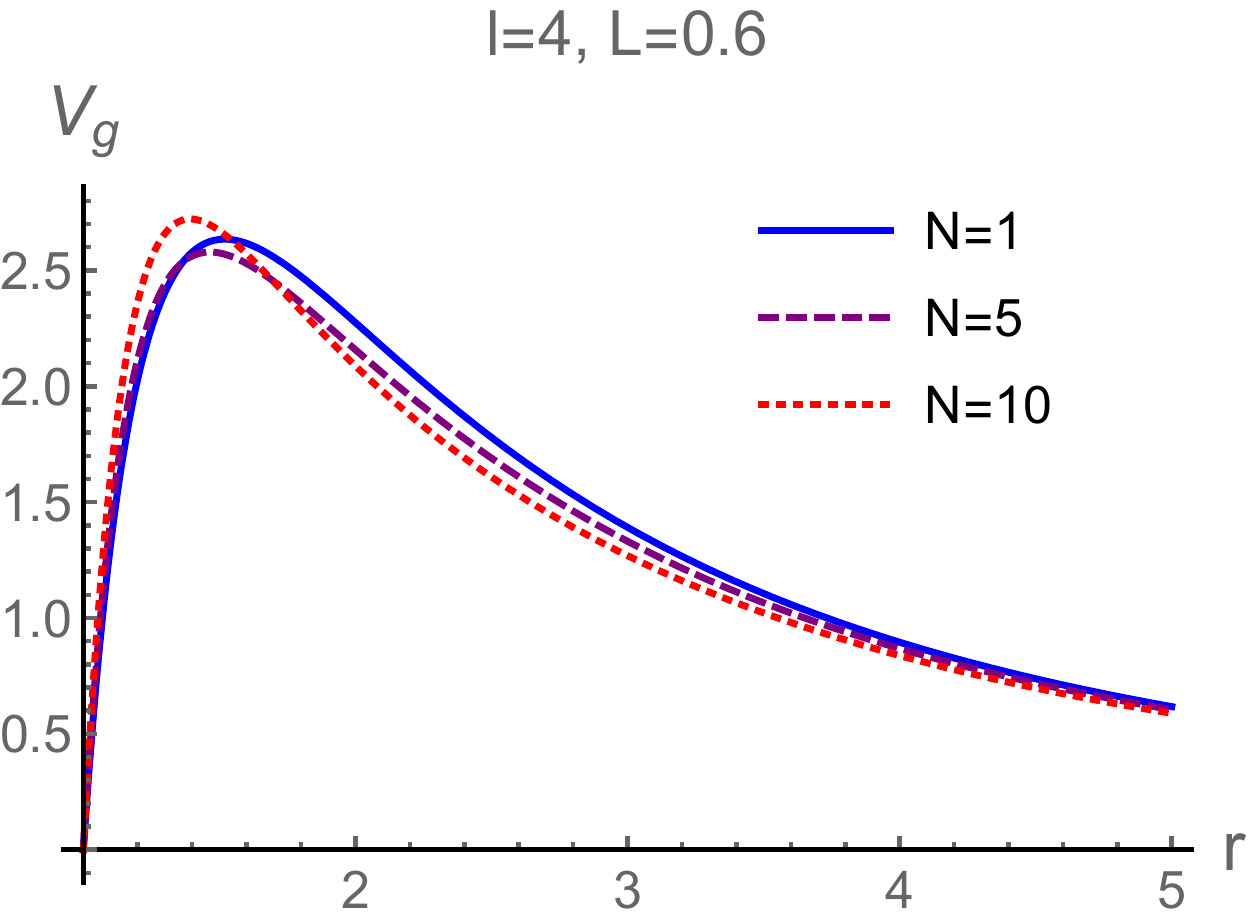}
\includegraphics[scale=0.46]{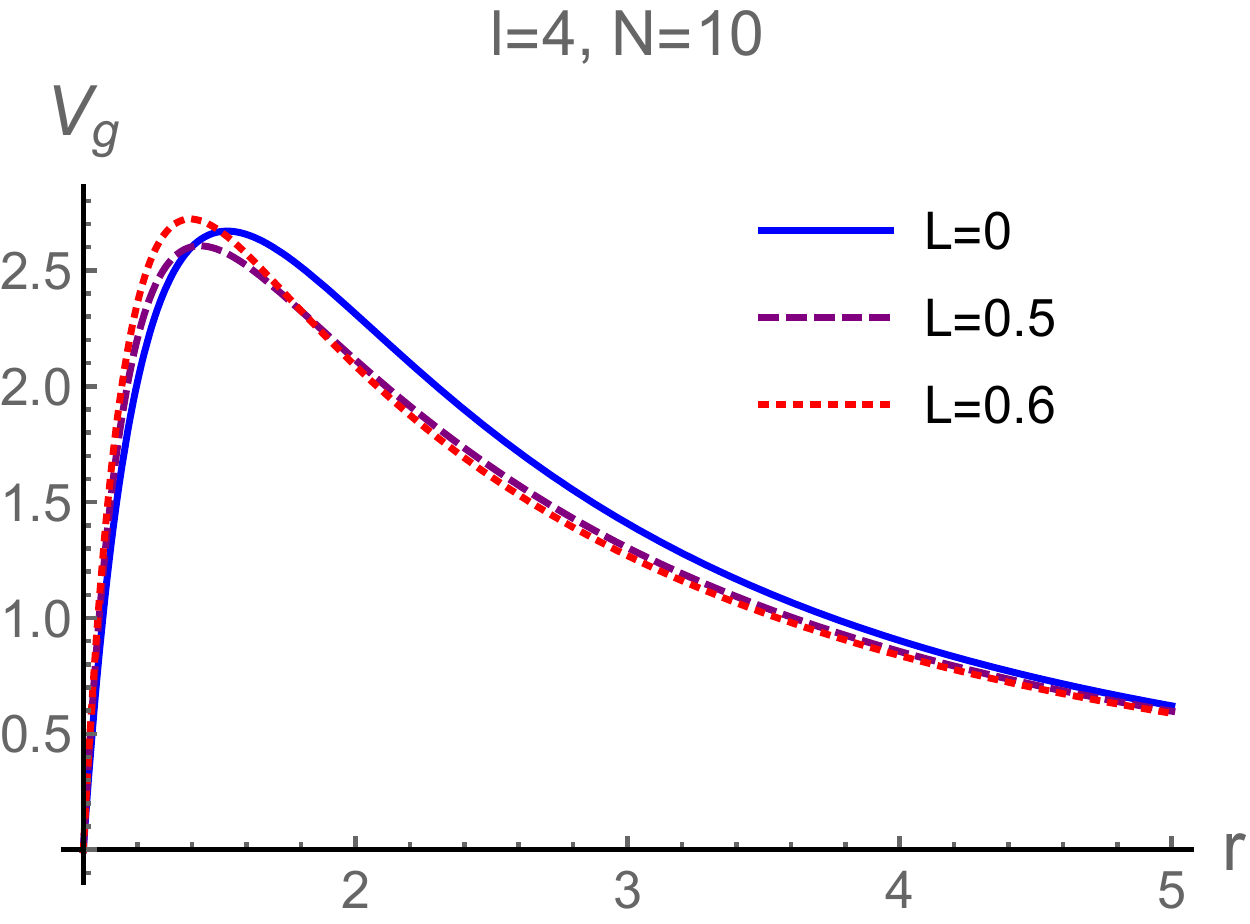}
\includegraphics[scale=0.46]{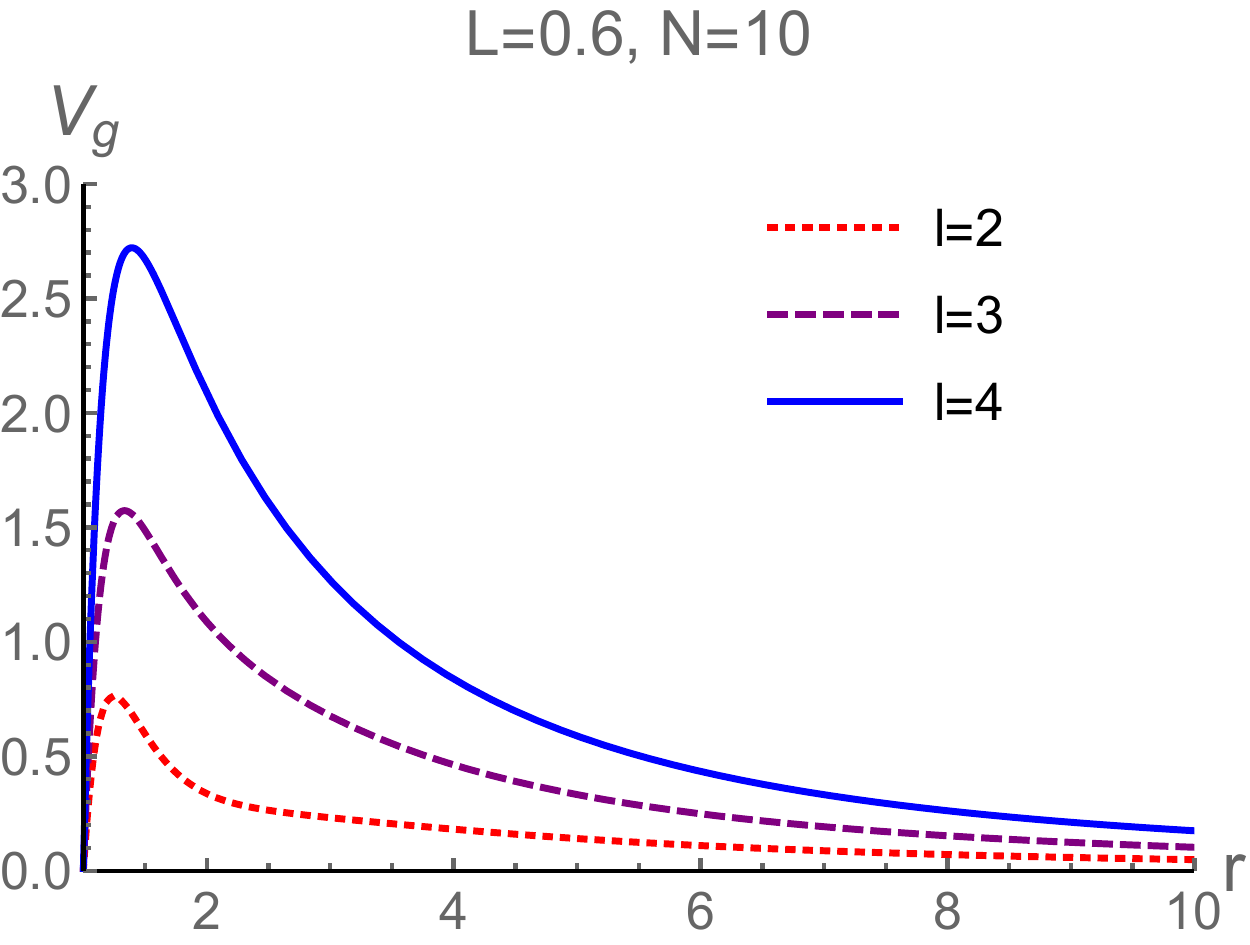}
\caption{The potential $V_g(r)$ given in Eq.~\eqref{potentialNN2} for different values of (from left to right) $N$, $L$, and the multipole number $l$. The values of the parameters are given in the figures.} 
\label{potentials}
\end{figure*}

The master equation of the axial perturbations reads (see appendix~\ref{deapp} for the derivation)
\begin{equation}
\frac{d^2H^{(-)}}{dr_*^2}+\omega^2H^{(-)}=V_g(r)H^{(-)}\,.\label{mastergggg}
\end{equation}
On the above equation, $\omega$ is the QNM frequency, $r_*$ is the tortoise radius defined by
\begin{equation}
\frac{dr}{dr_*}=f(r)\,,
\end{equation}
and the potential $V_g(r)$ is
\begin{equation}
V_g(r)=f(r)\left[\frac{l\left(l+1\right)}{r^2}-\frac{2}{r^2}-Z\frac{d}{dr}\left(\frac{f(r)\frac{dZ}{dr}}{Z^2}\right)\right]\,,\label{potentialgggg}
\end{equation}
where $Z\equiv \sqrt{S(r)}r$ and $l$ is the multipole number.

If we use the conformal factor given in Eq.~\eqref{confS}, the potential $V_g(r)$ can be written as
\begin{equation}
V_g(r)=f(r)\left[\frac{l\left(l+1\right)}{r^2}-\frac{3}{r^3}+F_1(r)N+F_2(r)N^2\right]\,,\label{potentialNN2}
\end{equation}
where 
\begin{align}
F_1(r)&=-\frac{2L^2\left(5r^3-6r^2+3L^2r-4L^2\right)}{r^3\left(r^2+L^2\right)^2}\,,\label{F1r}\\
F_2(r)&=\frac{4L^4\left(r-1\right)}{r^3\left(r^2+L^2\right)^2}\,.\label{F2r}
\end{align}
It can be easily seen that the master equation of the axial perturbations of the Schwarzschild black hole, that is, the Regge-Wheeler equation \cite{Regge:1957td}, is recovered when either $N=0$ or $L=0$. In Fig.~\ref{potentials}, we have shown the potential $V_g(r)$ given in Eq.~\eqref{potentialNN2} for different values of the parameters. According to these figures, it can be seen that increasing the values of either $N$ or $L$ in the range of $L<0.6$ would slightly decrease the height of the potential first, then increase it as $N$ and $L$ get larger. In addition, one can see that a larger multipole number $l$ significantly increases the height of the potential.

\section{QNM frequencies: the 6th order WKB method}\label{sect.wkb}
With the master equation \eqref{mastergggg} of the axial gravitational perturbations, the QNM frequencies can be calculated by treating the master equation as an eigenvalue problem with proper boundary conditions. Technically, there are various methods to calculate the QNMs, ranging from numerical approaches \cite{Leaver:1986gd,Jansen:2017oag} to semi-analytic methods (see Refs.~\cite{Nollert:1999ji,Berti:2009kk,Konoplya:2011qq,Berti:2015itd} and references therein).

Among the plethora of technical methods, in this paper we will use a semi-analytical approach, which is constructed on the WKB approximation, to evaluate the QNM frequencies. This method was firstly formulated in Ref.~\cite{Schutz:1985zz}. After that, the 1st order WKB method was extended to the 3rd and 6th order WKB approximation in Refs.~\cite{Iyer:1986np,Konoplya:2003ii}, respectively. Recently, a further extension of the WKB method up to the 13th order has been developed in Ref.~\cite{Matyjasek:2017psv}. With the WKB method, the QNM frequencies can be directly evaluated by using a simple formula as long as the potential term in the master equation is known. It should be highlighted that the WKB method is accurate when the multipole number $l$ is larger than the overtone $n$ \cite{Berti:2009kk}. Therefore, in the following discussions, we will devote to the QNMs of the fundamental modes $n=0$. We shall emphasize that for astrophysical black holes, the fundamental modes have the longest decay time and would dominate the late time signal during the ringdown stage.

The idea of the WKB method relies on the boundary conditions that we need to impose when calculating the QNM frequencies. At spatial infinity ($r_*\rightarrow\infty$), only outgoing waves moving away from the black hole exist. On the other hand, there can only exist ingoing waves moving toward the black hole at the event horizon ($r_*\rightarrow-\infty$) because nothing can escape from the event horizon. In order to encompass these boundary conditions, we treat the problem as a quantum scattering process without incident waves, while the reflected and the transmitted waves have comparable amounts of amplitudes. This can be achieved by assuming the peak value of the effective potential $V_{\textrm{eff}}(r_*)\equiv -\omega^2+V$ to be slightly larger than zero. There will be two classical turning points at the vicinity of the peak. At the regions far away from the turning points ($r_*\rightarrow\pm\infty$), we use the boundary conditions and solve the master equation with the help of the WKB approximation up to a desired order. Near the peak, the differential equation is solved by expanding the potential into a Taylor series up to a corresponding order. After matching the solution near the peak with those derived from the WKB approximation simultaneously at the two classical turning points, the numerical values of the QNM frequencies $\omega$ can be deduced according to the matching conditions.

In the 6th order WKB method, the QNM frequencies can be evaluated with the following formula \cite{Schutz:1985zz,Iyer:1986np,Konoplya:2003ii}
\begin{equation}
\frac{i\left(\omega^2-V_m\right)}{\sqrt{-2V_m''}}-\sum_{i=2}^6\Lambda_i=n+\frac{1}{2}\,,\label{WKBformal}
\end{equation}
where the index $m$ denotes the quantities evaluated at the peak of the potential. $V_m''$ is the second order derivative of the potential with respect to $r_*$, calculated at the peak. $\Lambda_i$ are constant coefficients resulting from higher order WKB corrections. These coefficients contain the value and derivatives (up to the 12th order) of the potential at the peak.{\footnote{The explicit expressions of $\Lambda_i$ are given in Refs.~\cite{Iyer:1986np,Konoplya:2003ii} (see Eqs.~(1.5a) and (1.5b) in Ref.~\cite{Iyer:1986np}, and the appendix in Ref.~\cite{Konoplya:2003ii}).}}

\subsection{Fundamental QNMs}
\begin{figure*}[t]
\includegraphics[scale=0.5]{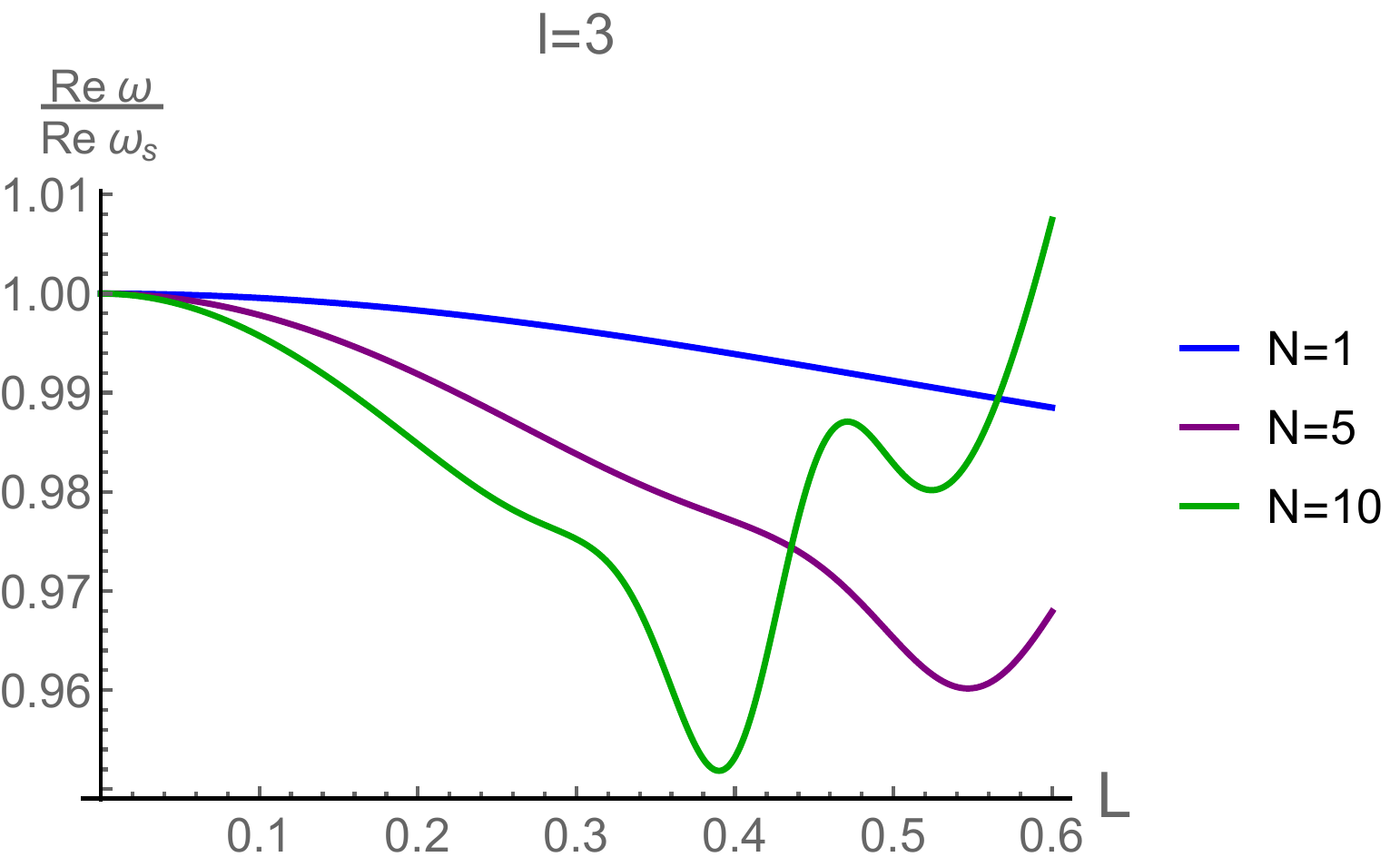}
\includegraphics[scale=0.5]{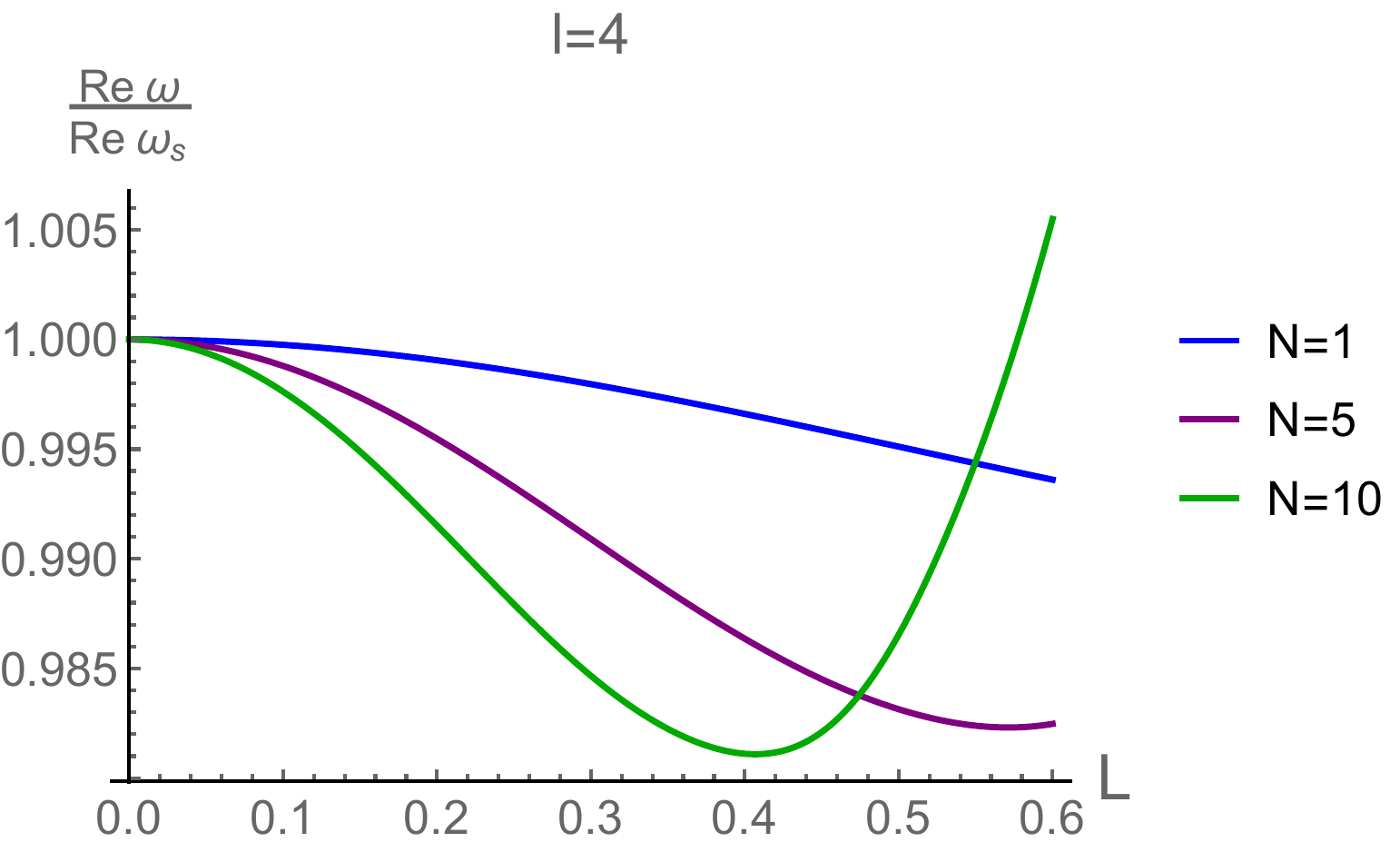}
\includegraphics[scale=0.5]{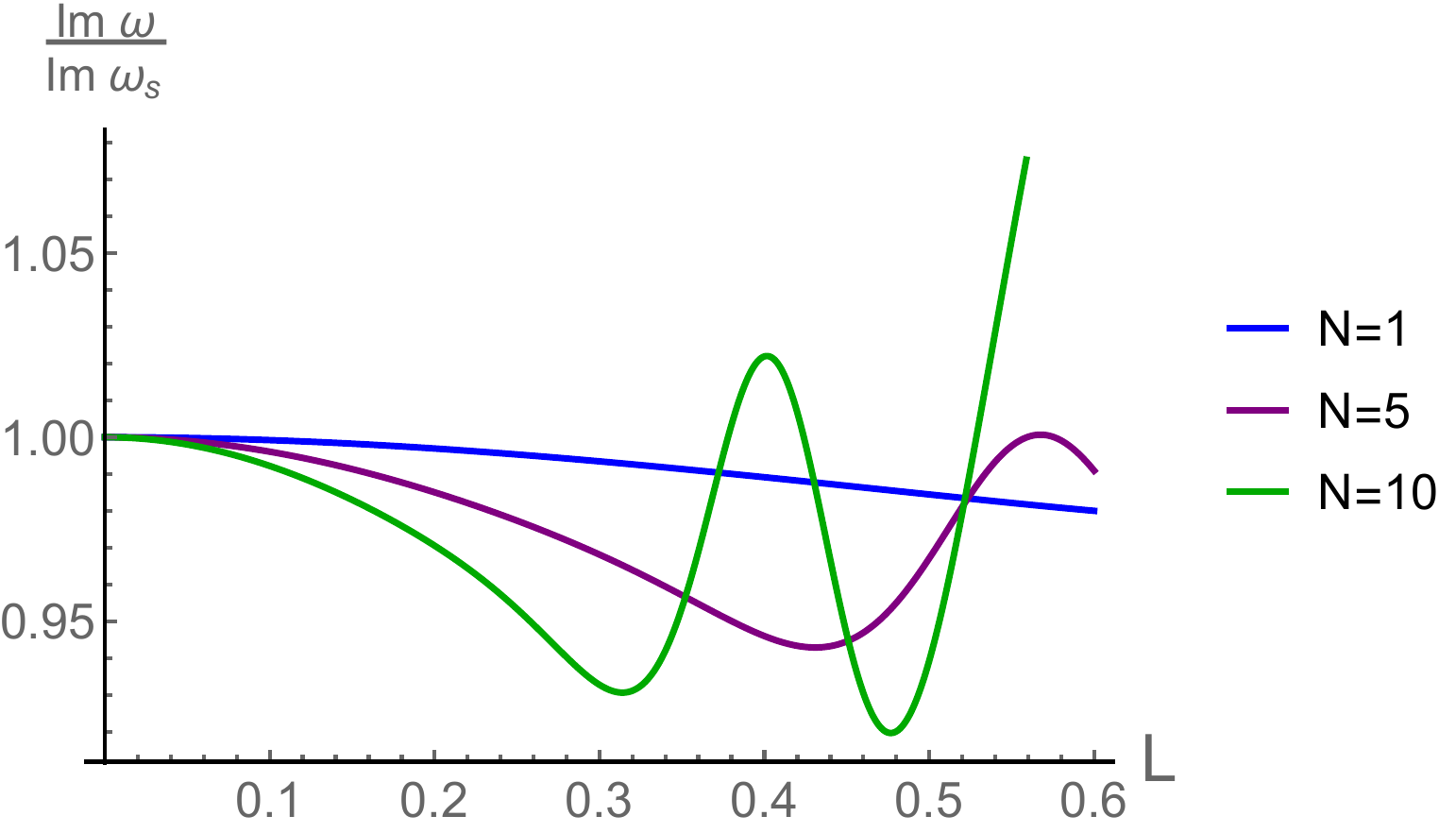}
\includegraphics[scale=0.5]{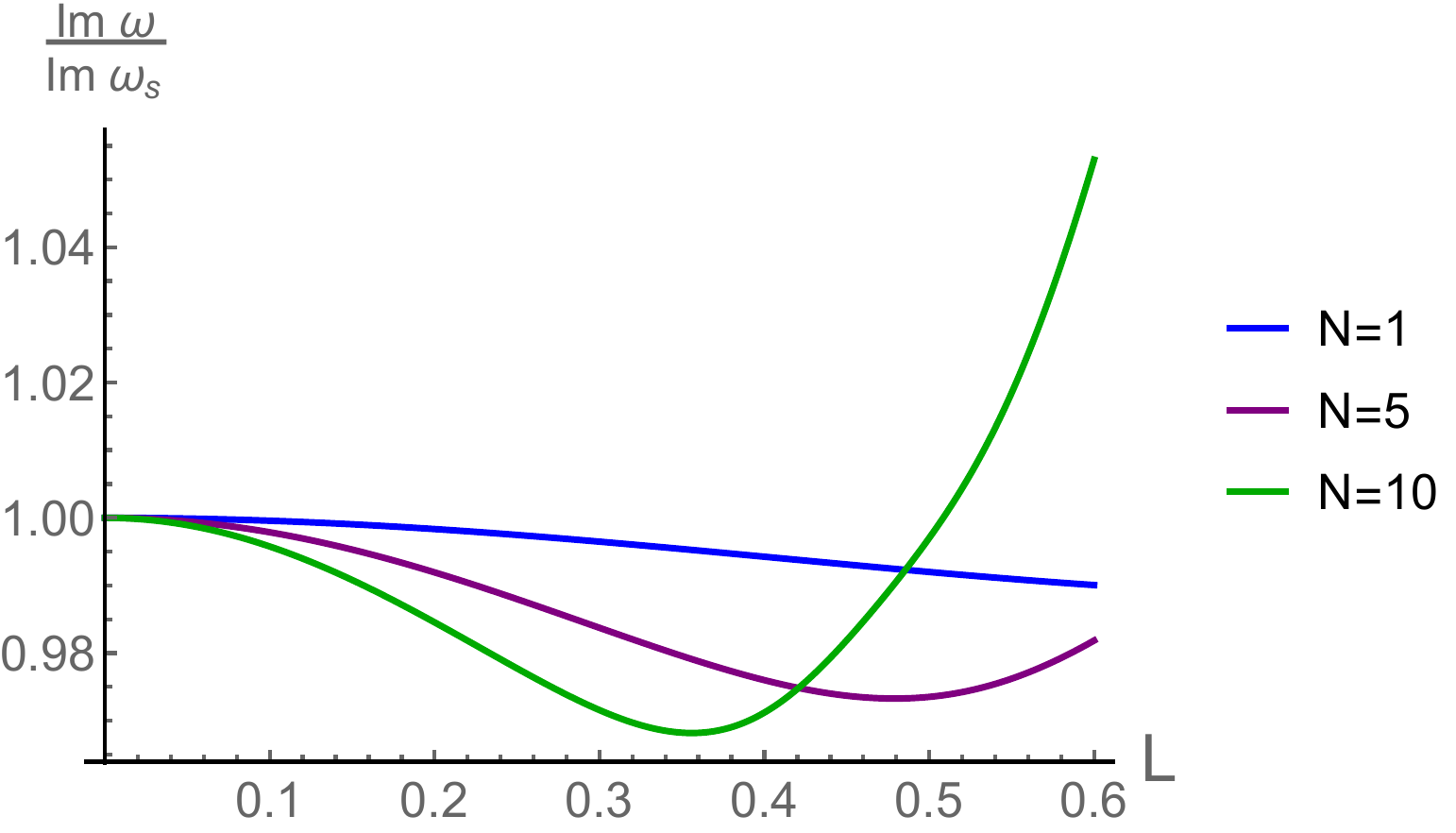}
\caption{\label{l3QNM}The real part (upper) and the imaginary part (lower) of the fundamental QNM frequencies of the non-singular black hole in conformal gravity are presented with respect to $L$. Different curves represent different values of $N$. The multipole number is chosen to be $l=3$ (left) and $l=4$ (right), respectively.}
\end{figure*}

In Fig.~\ref{l3QNM}, we calculate the real part (upper) and the imaginary part (lower) of the fundamental QNM frequencies of the non-singular black hole using the 6th order WKB formula. The frequencies are exhibited with respect to the parameter $L$.{\footnote{Note that the parameter $L$ has been constrained with X-ray observational data. For Kerr black hole, $L$ should be $L\le0.6$ when $N=2$ \cite{Bambi:2017yoz}. Recently, a more stringent constraint ($L\le0.225$ when $N=1$) has been derived in Ref.~\cite{Zhou:2018bxk}.}} Different curves correspond to different values of $N$. We consider $l=3$ (left) and $l=4$ (right) in this figure. 

From Fig.~\ref{l3QNM}, one can see that both the real part $\textrm{Re }\omega$ and the absolute value of the imaginary part $|\textrm{Im }\omega|$ of the frequencies would slightly decrease when $L$ starts to deviate from zero. When $L$ is getting larger (still within the parameter space of our interest, i.e., $L\le 0.6$), both $\textrm{Re }\omega$ and $|\textrm{Im }\omega|$ would increase. The overall tendency is more apparent when $N$ is larger. It should be noticed that for smaller value of $l$ (e.g. $l=3$), the frequencies possess a non-trivial oscillating behaviors when we increase the value of $L$. This is in contrary to the QNMs of the massless scalar field shown in Ref.~\cite{Toshmatov:2017bpx}, where the oscillating behaviors are absent. We will briefly compare the QNMs of the axial gravitational perturbations and those of the massless scalar field in subsection~\ref{comtwo}

\subsection{Eikonal QNMs}
In the eikonal limit where $l\rightarrow\infty$, the WKB method is accurate in calculating the QNM frequencies. In this limit, the potential $V_g(r)$ can be approximated as
\begin{equation}
V_g(r)\approx f(r)\frac{l^2}{r^2}\,,\label{eikonalpot}
\end{equation}
which is independent of the conformal factor. It can be seen that the QNM frequencies in large $l$ limit reduce to those of the Schwarzschild black hole. One cannot distinguish the non-singular black hole and the Schwarzschild black hole by using the QNMs in the eikonal limit.

Actually, the fact that the potential can be approximated as Eq.~\eqref{eikonalpot} gives rise to a straightforward way to calculate QNM frequencies. It can be shown that the peak of the potential \eqref{eikonalpot} locates exactly on the null circular orbit of the black hole. Therefore, the QNMs in the eikonal limit of a stationary, spherically symmetric, and asymptotically flat black hole can be obtained according to the properties of the null circular orbit \cite{Cardoso:2008bp}. More precisely, the QNM frequency in the eikonal limit can be deduced from \cite{Cardoso:2008bp}
\begin{equation}
\omega\approx\Omega_cl-i(n+1/2)|\lambda_c|\,,\label{eikonalGRfre}
\end{equation}
where $\Omega_c$ can be interpreted as the angular velocity of the null circular orbit and the parameter $\lambda_c$ stands for the Lyapunov exponent quantifying the instability of the orbit. 

One can see that the eikonal QNM frequencies of the non-singular black hole reduce to those of the Schwarzschild black hole, and they can be derived from the properties of the null circular orbit ($r=3/2$) via Eq.~\eqref{eikonalGRfre}. We would like to stress that in some other modified theories of gravity \cite{Konoplya:2017wot,Chen:2018vuw,Toshmatov:2018tyo}, the correspondence between the eikonal QNM frequencies and the properties of the null circular orbit around the black hole, i.e., Eq.~\eqref{eikonalGRfre}, could be violated. In those cases, it is possible to distinguish different theories by comparing their eikonal QNM frequencies.

\subsection{Large $L$ limit}
In this subsection, we intend to discuss the QNM frequencies when $L$ is large. It is expected that in this limit the deviations of the non-singular black hole from the Schwarzschild counterpart would be significant, jeopardizing the physical applicability of this region of parameter space. In fact, several constraints on $L$ have been derived according to X-ray observational data \cite{Bambi:2017yoz,Zhou:2018bxk}. Here, we study the QNMs in large $L$ limit simply due to its mathematical interest.

\begin{figure}
\includegraphics[scale=0.5]{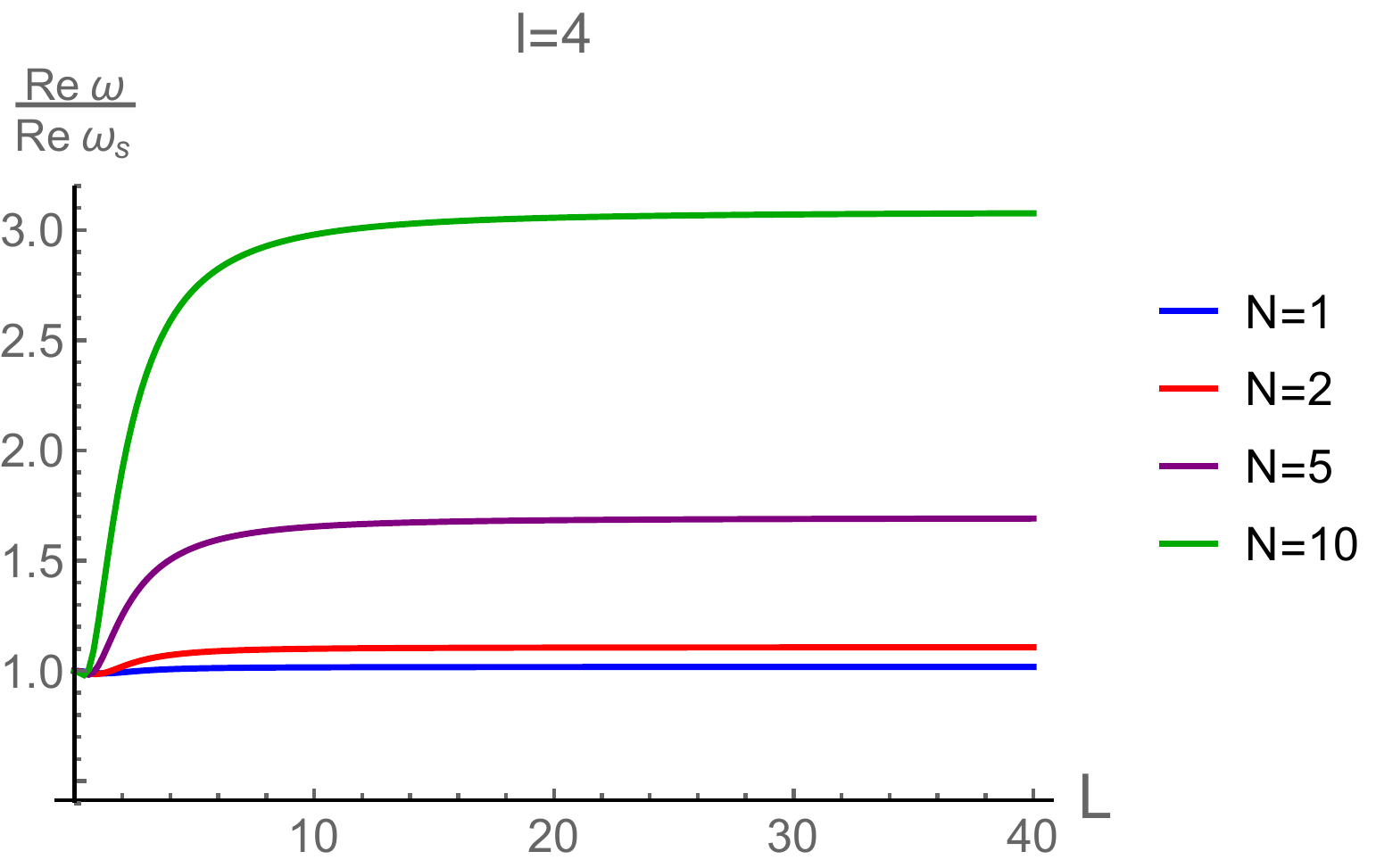}
\includegraphics[scale=0.5]{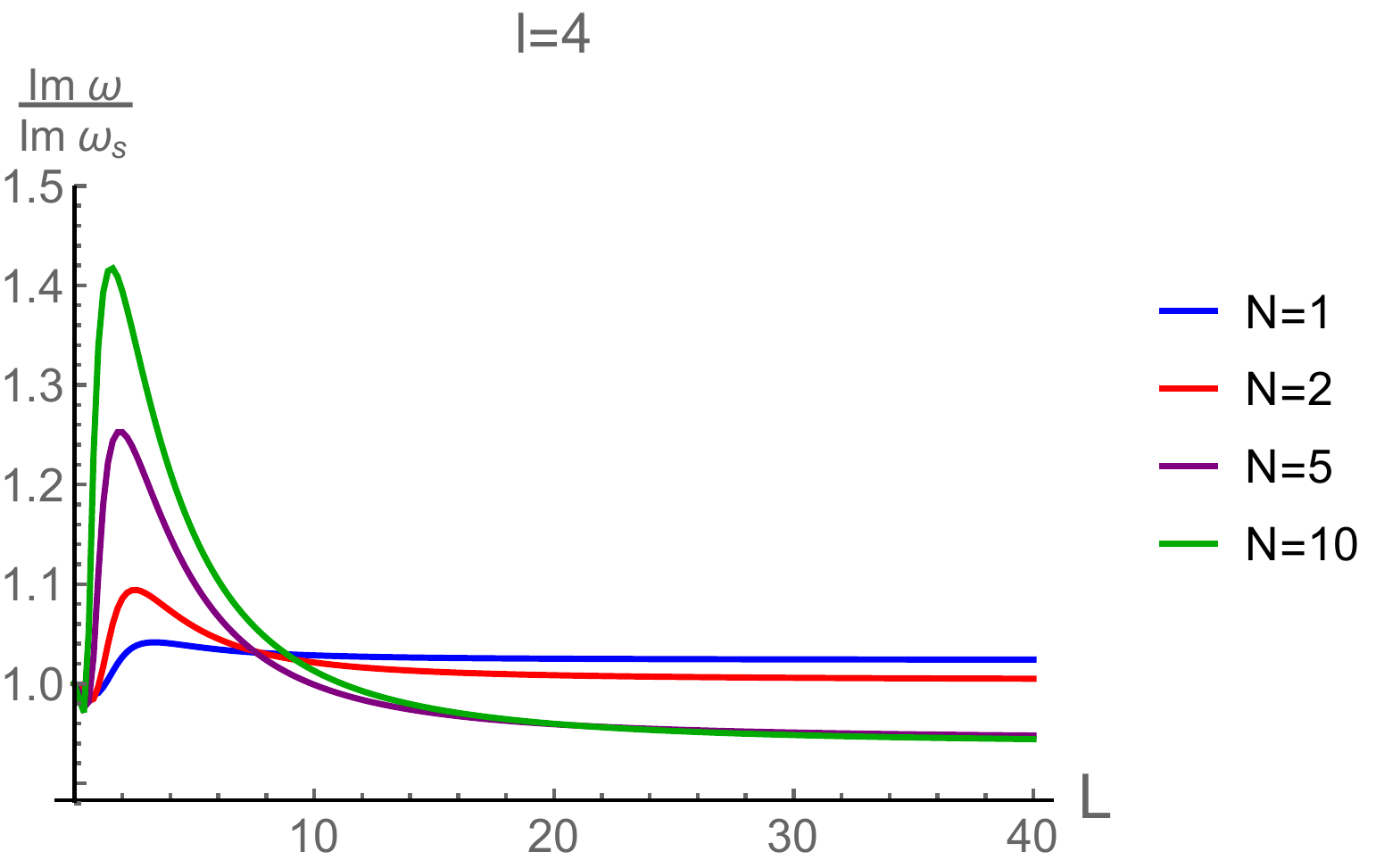}
\caption{\label{l4QNMlargeL}The upper (lower) figure shows the real (imaginary) part of the QNMs from the axial perturbations of the non-singular black hole in conformal gravity. The multipole number is chosen to be $l=4$. When $L$ is getting large, both the real and the imaginary parts of the frequencies approach a constant.}
\end{figure}

At large $L$ limit, the functions $F_1(r)$ and $F_2(r)$ in the potential \eqref{potentialNN2} can be approximated as
\begin{align}
F_1(r)&\approx\frac{8-6r}{r^3}\,,\\
F_2(r)&\approx\frac{4\left(r-1\right)}{r^3}\,.
\end{align}
It turns out that the potential $V_g$ is independent of $L$ and therefore the QNM frequencies approach a constant when $L$ becomes large (see Fig.~\ref{l4QNMlargeL}). More explicitly, the real part of the QNM frequencies increases with $L$. When $L$ is getting larger, $\textrm{Re }\omega$ approaches a constant value, which is determined by the parameter $N$. On the other hand, the absolute value of the imaginary part of the QNM frequencies increases and reaches its maximum value when $L\approx2$. When $L$ further increases, $|\textrm{Im }\omega|$ rapidly decreases and approaches a constant value, which is smaller if $N$ is larger.

\subsection{Large $N$ limit}
In Ref.~\cite{Toshmatov:2017kmw}, the authors assumed that the most natural criterion of selecting the preferred black hole solution from the huge family of conformally invariant solutions at the symmetry breaking is based on the amount of violation of the energy conditions. Nature might choose solutions which have less violation of the energy conditions. Consequently, the non-singular black hole solutions with a large value of $N$ are more preferred \cite{Toshmatov:2017kmw}. Therefore, it is necessary to study the QNMs at large $N$ limit and see whether the solutions in this limit are observationally preferred or not.

In large $N$ limit, the potential $V_g$ can be approximated as
\begin{equation}
V_g(r)\approx f(r)F_2(r)N^2\,.
\end{equation}
Using the 1st order WKB formula, the QNM frequencies in large $N$ limit can be deduced from the following equation
\begin{equation}
\omega\approx N\sqrt{\left(fF_2\right)_m}-i\left(n+\frac{1}{2}\right)\sqrt{\frac{-\left(fF_2\right)''_m}{2\left(fF_2\right)_m}}\,.\label{largeNQNM}
\end{equation}
where the index $m$ denotes the quantities evaluated at $r=r_m$, where $f(r)F_2(r)$ gets its maximum value. The prime stands for the derivative with respect to $r_*$, as in Eq.~\eqref{WKBformal}. 

According to the formula \eqref{largeNQNM} and the function $F_2(r)$ given in \eqref{F2r}, the real part of the frequency is linear in $N$. This can be seen from the green curves of the upper panel of Fig.~\ref{l4QNMlargeL}. On the other hand, the imaginary part of the QNM frequency in large $N$ limit is independent of $N$, while it is linear in the overtone $n$ (similar tendencies also appear for the QNMs of the massless scalar field perturbations \cite{Toshmatov:2017bpx}). One can see from the lower panel of Fig.~\ref{l4QNMlargeL} that the purple curves ($N=5$) and the green curves ($N=10$) are close to each other, especially at large $L$ limit where the imaginary part of the QNM frequency neither depends on $N$ nor $L$. 

According to our results, the QNM frequencies of non-singular black holes with a large $N$ seem to deviate significantly from their GR counterpart, i.e., the Schwarzschild black hole. Therefore, such non-singular black holes and the selection criterion based on the amount of violation of energy conditions proposed in Ref.~\cite{Toshmatov:2017kmw} might be observationally inconsistent.

Another important result according to Fig.~\ref{l4QNMlargeL} is that the imaginary part of the frequency does not change sign when the values of $N$ and $L$ are changed. As we have mentioned, when $N$ and $L$ get large, the imaginary part of the QNM frequency approaches a constant which is independent of the values of $N$ and $L$. Therefore, the sign of the imaginary part of the frequency remains unchanged, indicating that the perturbations are always damping oscillations and the black holes are stable against the axial perturbations. Actually, it is well known that the perturbations are stable as long as the potential $V_g$ is positive definite everywhere outside the horizon. The potential of our interest is indeed the case (see Fig.~\ref{potentials}).

\subsection{Time evolution and late-time tails}
\begin{figure*}[tt]
\centering
\graphicspath{{fig/}}
\includegraphics[scale=0.46]{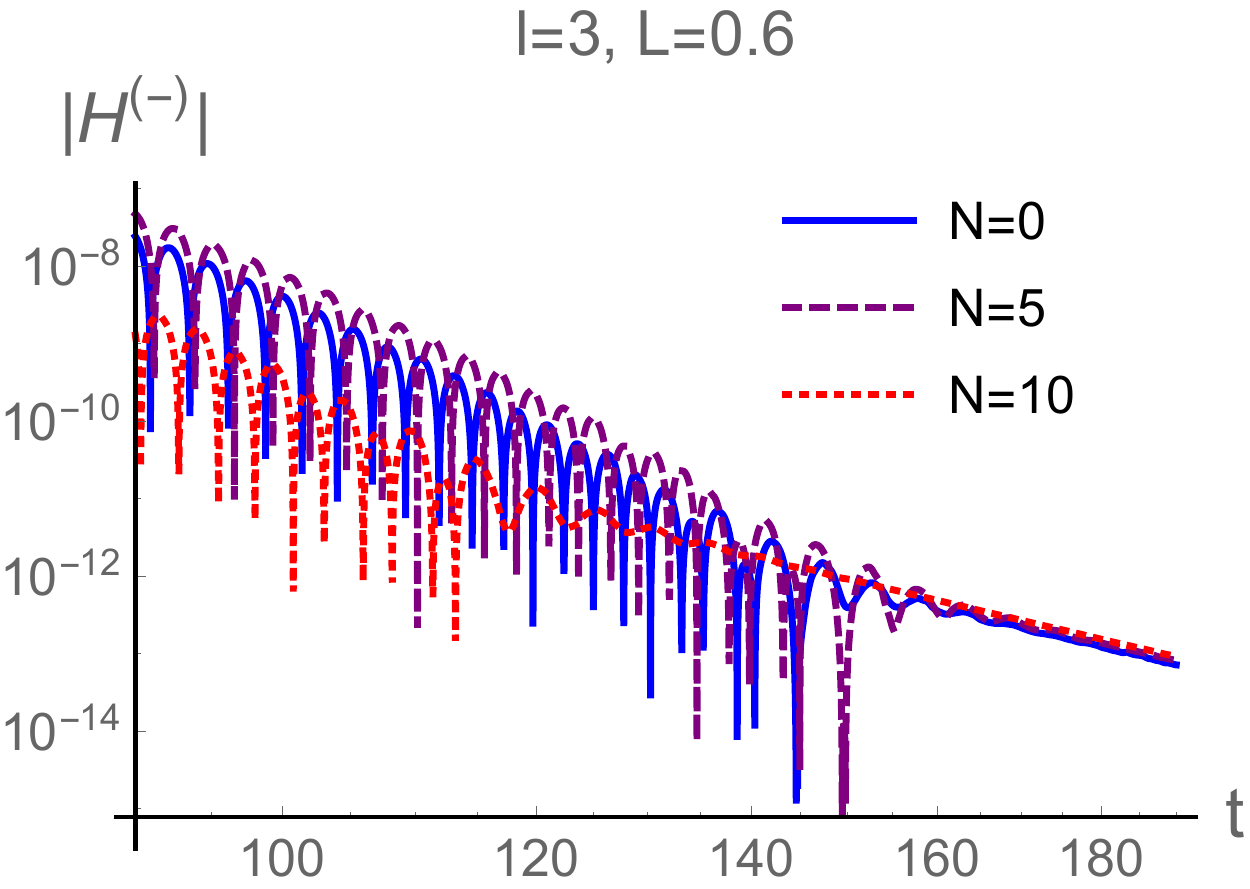}
\includegraphics[scale=0.46]{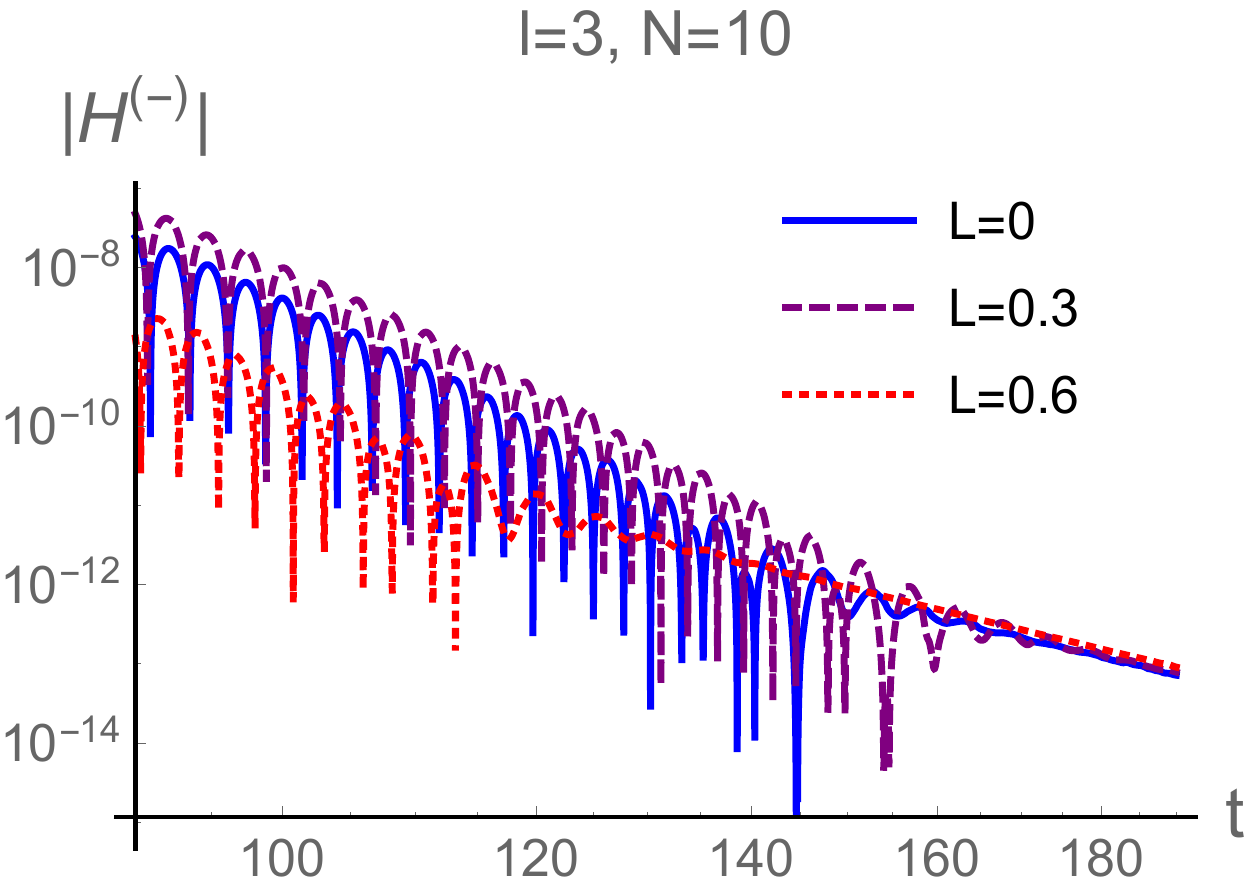}
\includegraphics[scale=0.46]{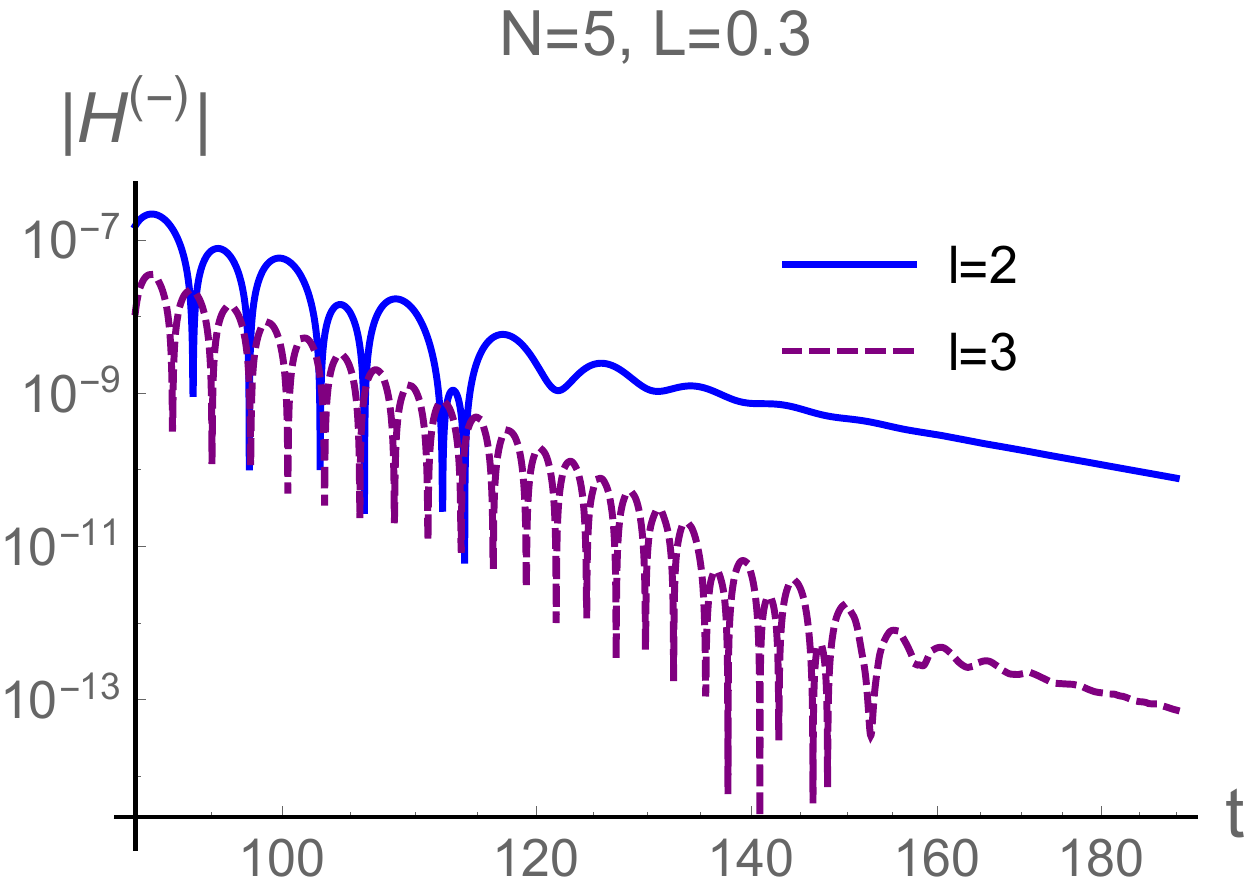}
\caption{The time domain profiles of the axial gravitational perturbations for the non-singular black holes in the conformal gravity with different values of (from left to right) $N$, $L$, and the multipole number $l$. The values of the parameters are given in the figures.} 
\label{timedomain}
\end{figure*}

In this subsection, we will study the time domain evolution of the axial gravitational perturbations as well as the late-time tail which is the power-law falloff of the perturbations. To derive the time evolution of the perturbations, we follow the standard strategy illustrated in Refs.~\cite{Gundlach:1993tp,Chirenti:2007mk,Aneesh:2018hlp}. More precisely, we rewrite the master equation \eqref{mastergggg} in the light cone coordinates as follows
\begin{equation}
\left[4\frac{\partial^2}{\partial u\partial v}+V_g(u,v)\right]H^{(-)}(u,v)=0\,,
\label{uvmaster}
\end{equation} 
where $u=t-r_*$ and $v=t+r_*$. This equation can be directly integrated numerically after appropriate initial data on $u=u_0$ and $v=v_0$ are imposed. The time domain profiles of the axial perturbations for different values of the parameters are shown in Fig.~\ref{timedomain}. It can be seen from the left and middle panels of Fig.~\ref{timedomain} that the tails are parallel to each other for different values of $N$ and $L$. Actually, the power law tail only depends on the multipole number $l$ and one can see from the right panel of Fig.~\ref{timedomain} that the perturbations with higher $l$ would decay faster.

\subsection{Comparison with the QNMs from massless scalar field}\label{comtwo}
Before closing this section, we would like to compare the QNMs of the axial gravitational perturbations with those from the massless scalar field perturbations. Unlike the QNMs of the electromagnetic perturbations which remain unchanged after conformal transformations \cite{Toshmatov:2017bpx}, both the scalar perturbations and the axial gravitational perturbations deviate from their Schwarzschild counterpart. In the previous discussion, we have already shown how the QNMs of the axial gravitational perturbations change in the presence of the additional conformal factor. In this subsection, we will briefly review the QNMs of the massless scalar field perturbations, which was studied in Ref.~\cite{Toshmatov:2017bpx}.

The master equation describing the massless scalar field perturbations is deduced from the Klein-Gordon equation, and it can be written as \cite{Toshmatov:2017bpx}:
\begin{equation}
\frac{d^2\Psi}{dr_*^2}+\omega^2\Psi=V_s(r)\Psi\,,
\end{equation}
where 
\begin{equation}
V_s(r)=f(r)\left[\frac{l\left(l+1\right)}{r^2}+\frac{1}{Z}\frac{d}{dr}\left(f(r)\frac{dZ}{dr}\right)\right]\,,
\end{equation}
where $Z=\sqrt{S(r)}r$ and the potential depends on the conformal factor $S(r)$.

\begin{figure}
\includegraphics[scale=0.5]{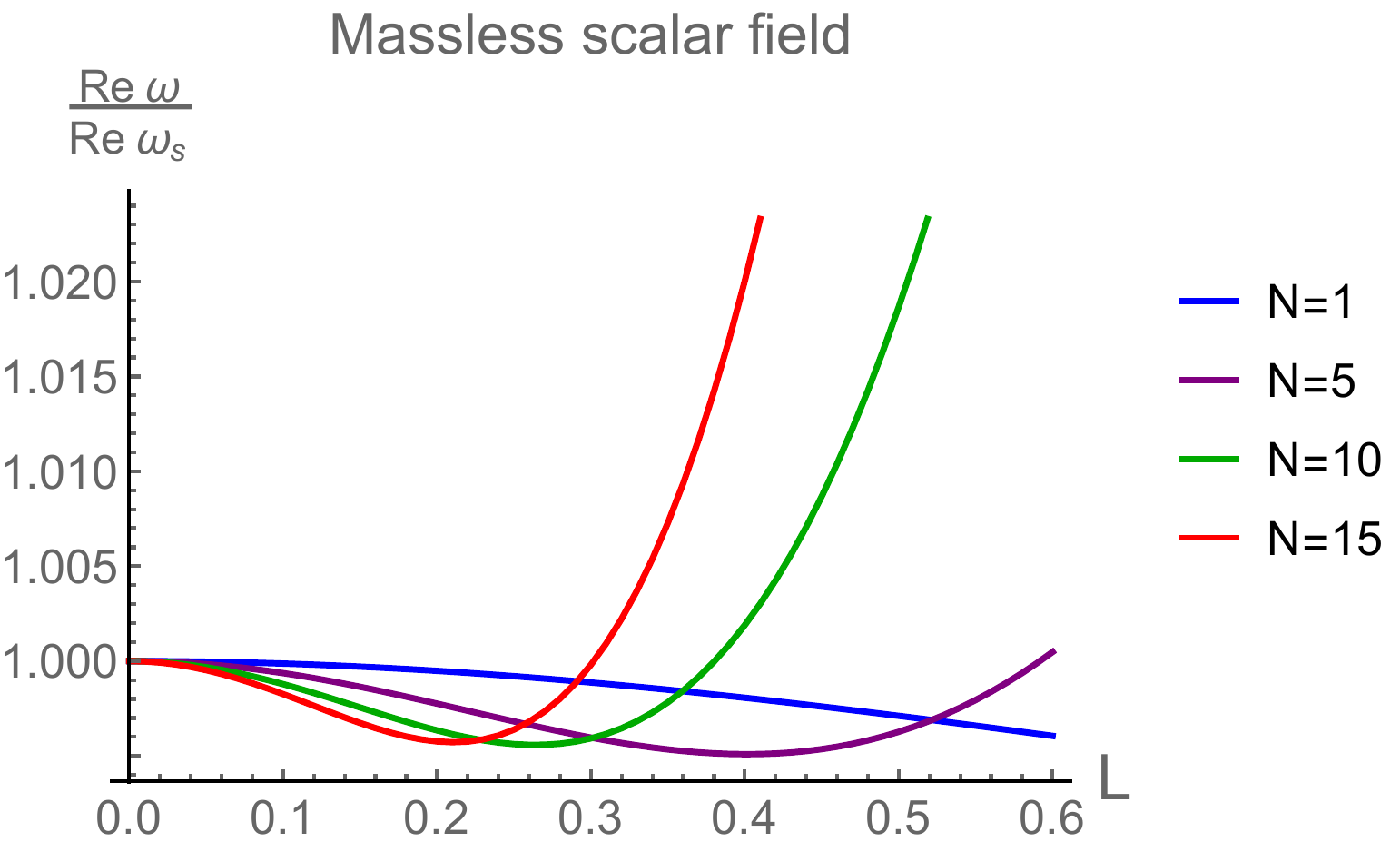}
\includegraphics[scale=0.5]{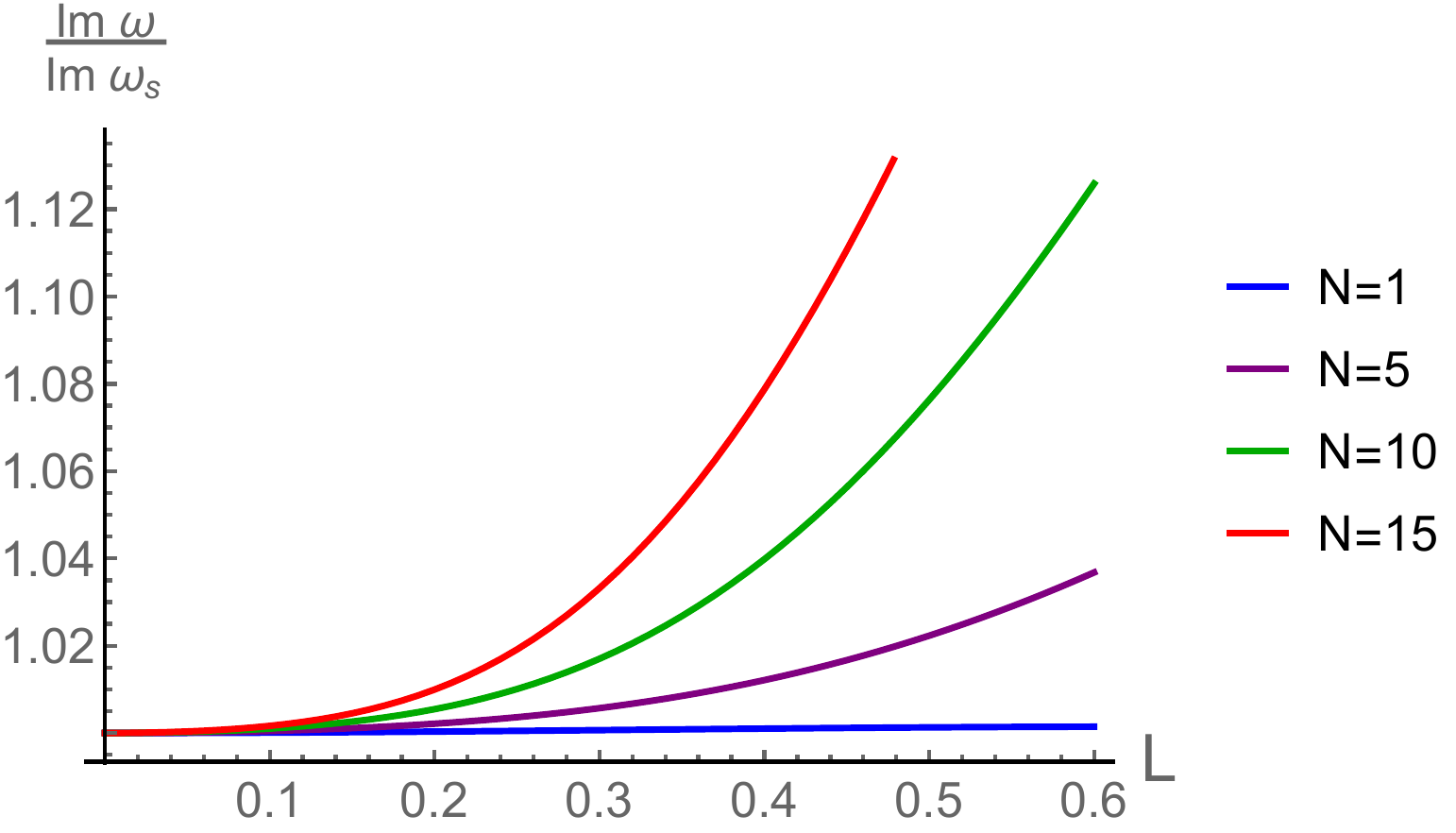}
\caption{\label{scalarQNM}The upper (lower) figure shows the real (imaginary) part of the QNMs from the massless scalar field of the non-singular black hole in conformal gravity with respect to $L$. The multipole number is chosen to be $l=3$. The results with $l=2$ have been given in Ref.~\cite{Toshmatov:2017bpx}.}
\end{figure}

Considering the conformal factor of our interest \eqref{confS}, the QNM frequencies of the massless scalar field perturbations are calculated by using the 6th order WKB method, as what we have done for the axial perturbations. The results are shown in Fig.~\ref{scalarQNM}. Here we choose the multipole number $l=3$. The results with $l=2$ were presented in Ref.~\cite{Toshmatov:2017bpx}. One can see that the real part of the frequencies $\textrm{Re }\omega$ would slightly decrease when we increase $L$ a little bit from zero. Then it increases with $L$. As for the imaginary part, it can be seen that $|\textrm{Im }\omega|$ increases with $L$ and $N$. This is different from what we have found in the axial perturbations. For the axial perturbations, both $\textrm{Re }\omega$ and $|\textrm{Im }\omega|$ would in general decrease then increase with $L$. Furthermore, for $l=3$ the axial QNM frequencies with a large $N$ possess a non-trivial oscillating behavior when we change the value of $L$.

\section{Conclusions}\label{conclu}

In this paper, we investigate the QNMs of the axial perturbations of the non-singular black hole, which can be a solution within a family of conformal gravity theories. In conformal gravity, the spacetime respects the conformal symmetry and all the metrics transformed conformally are physically equivalent. It is straightforward to remove the spacetime singularity by introducing a conformal factor such that the spacetime described by the new metric $\hat{g}_{\mu\nu}$ is everywhere non-singular. The spacetime singularity is just a mathematical artifact of choosing different conformal gauges. The non-singular spacetime is geodesically complete and the curvature invariants (in the notion of coordinate transformations) turn out to be finite everywhere in the spacetime.

In order to recover the present universe that we are living in, there must be a phase transition where the conformal symmetry is broken. After the phase transition, different conformally related spacetime metrics give rise to different observational consequences. The exact expressions of the master equation of the gravitational perturbations should be derived from the perturbed gravitational equations and should therefore depend on the underlying gravitational theory. In the absence of a specific conformal gravity under consideration, we derive the master equation by assuming that the spacetime is described by the Einstein equation coupled with the effective energy momentum tensor of an anisotropic fluid. The results obtained in this paper can be regarded as a generic feature of a family of conformal gravity from a phenomenological perspective. 

As pointed out in this paper, the QNMs of the axial gravitational perturbations are able to see the effects of conformal factors, like the cases of the massless scalar field QNMs studied in Ref.~\cite{Toshmatov:2017bpx}. However, for the QNMs of the massless scalar field perturbations, $\textrm{Re }\omega$ tends to decrease slightly when $L$ slightly increases from zero, then it starts to increase with $L$.  The absolute value of the imaginary part $|\textrm{Im }\omega|$ increases with $L$ and $N$. These behaviors have been presented in Fig.~\ref{scalarQNM} and also in the paper \cite{Toshmatov:2017bpx}. On the other hand, for the QNMs of the axial gravitational perturbations investigated in this paper, both $\textrm{Re }\omega$ and $|\textrm{Im }\omega|$ would decrease first with $L$, and then increase when $L$ is of order one (See Fig.~\ref{l3QNM}). Also, we have found a non-trivial oscillating behavior of the QNM frequencies as a function of $L$ when the multipole number is $l=3$. On the other hand, we have found that the non-singular black holes in conformal gravity cannot be distinguished from the Schwarzschild black hole by using the eikonal QNMs because the master equation becomes independent of the conformal factors. Finally, we have concluded that the parameters $L$ and $N$ for a non-singular black hole cannot be too large otherwise the QNM frequencies would deviate too much from those of the Schwarzschild black hole. This implies that the selection criterion based on the amount of violation of energy conditions, which was proposed in Ref.~\cite{Toshmatov:2017kmw}, is likely to be observationally disfavored. 

We would like to stress that the analysis done in this paper is only for non-rotating black holes. In reality, black holes should have spins and can be well-described by the Kerr metric in the context of GR. Testing rotating black holes in modified theories of gravity is extremely difficult. Actually, even the derivation of an exact solution of rotating black holes with an arbitrary spin can be extremely challenging due to the complexity of the gravitational equations. However, the class of rotating black holes in the conformal gravity proposed in Ref.~\cite{Bambi:2016wdn} exquisitely skirt this technical difficulty. The non-singular rotating black holes can be obtained by conformally transforming the Kerr metric. One can then apply a similar strategy to compare the QNM spectra of these rotating black holes with those of the Kerr black hole. We shall leave this interesting issue for a coming work.
 
\appendix 

\section{The derivation of the master Eq.~\eqref{mastergggg}}\label{deapp}
In order to study the QNMs of a static and spherically symmetric black hole, we consider a perturbed spacetime which is described by a non-stationary and axisymmetric metric as follows \cite{Chandrabook}:
\begin{align}
ds^2=&-e^{2\nu}\left(dx^0\right)^2+e^{2\psi}\left(dx^1-\sigma dx^0-q_2dx^2-q_3dx^3\right)^2\nonumber\\&+e^{2\mu_2}\left(dx^2\right)^2+e^{2\mu_3}\left(dx^3\right)^2\,,\label{metricg}
\end{align}
where $\nu$, $\psi$, $\mu_2$, $\mu_3$, $\sigma$, $q_2$, and $q_3$ are functions of time $t$ ($t=x^0$), radial coordinate $r$ ($r=x^2$), and polar angle $\theta$ ($\theta=x^3$). Since the system is axisymmetric, the metric functions can be assumed to be independent of the azimuthal angle $\phi$ ($\phi=x^1$). In the following derivation, the notation used in Ref.~\cite{Chandrabook} is strictly followed. The only difference is that the metric function $\omega$ used in Ref.~\cite{Chandrabook} is replaced with $\sigma$ in Eq.~\eqref{metricg} as we will use $\omega$ to denote the frequency of the perturbations. Note that $q_2$, $q_3$, and $\sigma$ are zero for a static and spherically symmetric spacetime. Therefore, when linearizing the field equations, these metric functions should be regarded as linear order perturbation quantities.

To proceed, we will use the tetrad formalism in which one defines a tetrad basis corresponding to the metric \eqref{metricg} (see Ref.~\cite{Chandrabook} for a clear introduction of the tetrad formalism):
\begin{align}
e^{\mu}_{(0)}&=\left(e^{-\nu},\quad\sigma e^{-\nu},\quad0,\quad0\right)\,,\nonumber\\
e^{\mu}_{(1)}&=\left(0,\quad e^{-\psi},\quad 0,\quad0\right)\,,\nonumber\\
e^{\mu}_{(2)}&=\left(0,\quad q_2e^{-\mu_2},\quad e^{-\mu_2},\quad0\right)\,,\nonumber\\
e^{\mu}_{(3)}&=\left(0,\quad q_3e^{-\mu_3},\quad 0,\quad e^{-\mu_3}\right)\,,\label{tetradbasis111}
\end{align}
where the tetrad indices are enclosed in parentheses to distinguish them from the tensor indices. Essentially, in the tetrad formalism all the relevant quantities defined on the coordinate basis of $\hat{g}_{\mu\nu}$ are projected onto a specific basis of $\eta_{(a)(b)}$ by using the corresponding tetrad basis. Usually, it is convenient to assume $\eta_{(a)(b)}$ to be the Minkowskian metric.
Upon this construction, any vector or tensor field can be projected onto the tetrad frame in which the field is expressed through its tetrad components:
\begin{align}
A_{\mu}&=e_{\mu}^{(a)}A_{(a)}\,,\quad A_{(a)}=e_{(a)}^{\mu}A_{\mu}\,,\nonumber\\
B_{\mu\nu}&=e_{\mu}^{(a)}e_{\nu}^{(b)}B_{(a)(b)}\,,\quad B_{(a)(b)}=e_{(a)}^{\mu}e_{(b)}^{\nu}B_{\mu\nu}\,.
\end{align}

\subsection{Perturbed energy momentum tensor}
In the tetrad frame, the perturbed energy-momentum tensor of an anisotropic fluid reads
\begin{align}
\delta T_{(a)(b)}=&\,(\rho+p_2)\delta(u_{(a)}u_{(b)})+(\delta\rho+\delta p_2)u_{(a)}u_{(b)}\nonumber\\
&+(p_1-p_2)\delta(x_{(a)}x_{(b)})+(\delta p_1-\delta p_2)x_{(a)}x_{(b)}\nonumber\\&+\delta p_2\eta_{(a)(b)}.
\end{align}
After considering the constraints on $u^\mu$ and $x^\mu$, that is, Eq.~\eqref{fourvelocity} and $u^\mu x_\mu=0$, we find that the axial components of the perturbed energy momentum tensor in the tetrad frame vanish:
\begin{align}
\delta T_{(1)(0)}&=\delta T_{(1)(2)}=\delta T_{(1)(3)}=0\,.
\end{align}

\subsection{Perturbed Einstein equation}
In the tetrad frame, the Einstein equation can be rewritten as
\begin{equation}
R_{(a)(b)}-\frac{1}{2}\eta_{(a)(b)}R=8\pi T_{(a)(b)}\,.\label{eineq}
\end{equation}
Since the axial components of the perturbed energy momentum tensor vanish, the master equation of the axial perturbations can be derived from $R_{(a)(b)}|_{\textrm{axial}}=0$. We then consider the $(1,3)$ and $(1,2)$ components of $R_{(a)(b)}|_{\textrm{axial}}=0$ and get:
\begin{align}
&\left[Sr^2e^{\nu-\mu_2}\left(q_{2,3}-q_{3,2}\right)\right]_{,2}-Sr^2e^{-\nu+\mu_2}\left(\sigma_{,3}-q_{3,0}\right)_{,0}\nonumber\\=&\,0\,,\label{35}\\
&\left[Sr^2e^{\nu-\mu_2}\left(q_{3,2}-q_{2,3}\right)\sin^3{\theta}\right]_{,3}\nonumber\\&-\,S^2r^4e^{-\nu-\mu_2}\left(\sigma_{,2}-q_{2,0}\right)_{,0}\sin^3{\theta}=0\,.\label{36}
\end{align}
Then, we define
\begin{equation}
Q\equiv Sr^2e^{\nu-\mu_2}\left(q_{2,3}-q_{3,2}\right)\sin^3{\theta}\,,
\end{equation}
with which Eqs.~\eqref{35} and \eqref{36} can be rewritten as
\begin{align}
e^{\nu-\mu_2}\frac{Q_{,2}}{Sr^2\sin^3{\theta}}&=\left(\sigma_{,3}-q_{3,0}\right)_{,0}\,,\label{4022}\\
e^{\nu+\mu_2}\frac{Q_{,3}}{S^2r^4\sin^3{\theta}}&=-\left(\sigma_{,2}-q_{2,0}\right)_{,0}\,.\label{40}
\end{align}
By differentiating Eqs.~\eqref{4022} and \eqref{40} and eliminating $\sigma$, we obtain
\begin{align}
&\frac{1}{\sin^3{\theta}}\left(\frac{e^{\nu-\mu_2}}{Sr^2}Q_{,2}\right)_{,2}+\frac{e^{\nu+\mu_2}}{S^2r^4}\left(\frac{Q_{,3}}{\sin^3{\theta}}\right)_{,3}\nonumber\\
=&\,\frac{Q_{,00}}{Sr^2\sin^3{\theta}e^{\nu-\mu_2}}\,.\label{41w}
\end{align}

\subsection{The master equation}
To derive the master equation, we consider the ansatz \cite{Chandrabook}
\begin{equation}
Q(r,\theta)=Q(r)Y(\theta)\,,\qquad B(r,\theta)=B(r)Y_{,\theta}/\sin{\theta}\,,\label{43}
\end{equation}
where $Y(\theta)$ is the Gegenbauer function satisfying \cite{Abramow}
\begin{equation}
\frac{d}{d\theta}\left(\frac{1}{\sin^3{\theta}}\frac{dY}{d\theta}\right)=-\mu^2\frac{Y}{\sin^3{\theta}}\,,\label{gegen1}
\end{equation}
where $\mu^2=(l-1)(l+2)$. Eq.~\eqref{41w} can be rewritten as
\begin{align}
\left(\frac{e^{\nu-\mu_2}}{Sr^2}Q_{,r}\right)_{,r}+\left(\frac{\omega^2}{Sr^2e^{\nu-\mu_2}}-\frac{e^{\nu+\mu_2}\mu^2}{S^2r^4}\right)Q=0\,,\label{461}
\end{align}
where we have used the Fourier decomposition $\partial_t\rightarrow-i\omega$.

We introduce the following definitions
\begin{equation}
H^{(-)}\equiv \frac{Q}{Z}\,,\label{subfr1}
\end{equation}
where $Z\equiv \sqrt{S}r$, and consider the tortoise radius $r_*$ which satisfies
\begin{equation}
\frac{dr}{dr_*}=e^{\nu-\mu_2}=f(r)\,.\label{subfr2}
\end{equation}
The master equation \eqref{461} becomes
\begin{equation}
\frac{d^2H^{(-)}}{dr_*^2}+\omega^2H^{(-)}=\,\left[-Z\left(\frac{Z_{,r_*}}{Z^2}\right)_{,r_*}+\frac{e^{2\nu}\mu^2}{Z^2}\right]H^{(-)}\,.\label{fr111}
\end{equation}
Since $Z=\sqrt{S}r$ and $e^{2\nu}=Sf$, Eq.~\eqref{fr111} can be rewritten  as
\begin{align}
\frac{d^2H^{(-)}}{dr_*^2}+\omega^2H^{(-)}=&\,f(r)\left[\frac{\mu^2}{r^2}-Z\frac{d}{dr}\left(\frac{f(r)\frac{dZ}{dr}}{Z^2}\right)\right]H^{(-)}\nonumber\\
=&\,V_g(r)\,,
\end{align}
completing the derivation of Eq.~\eqref{mastergggg}.

\acknowledgments

CYC would like to thank R. A. Konoplya for providing the WKB approximation. CYC and PC are supported by Taiwan National Science Council under Project No. NSC 97-2112-M-002-026-MY3, Leung Center for Cosmology and Particle Astrophysics (LeCosPA) of National Taiwan University, and Taiwan National Center for Theoretical Sciences (NCTS). PC is in addition supported by US Department of Energy under Contract No. DE-AC03-76SF00515.


\begin{thebibliography}{99}

\bibitem{Abbott:2016blz}
  B.~P.~Abbott {\it et al.} [LIGO Scientific and Virgo Collaborations],
  Phys.\ Rev.\ Lett.\  {\bf 116} (2016) no.6,  061102.
  
\bibitem{Abbott:2017oio}
  B.~P.~Abbott {\it et al.} [LIGO Scientific and Virgo Collaborations],
  Phys.\ Rev.\ Lett.\  {\bf 119} (2017) no.14,  141101.

\bibitem{TheLIGOScientific:2017qsa}
  B.~P.~Abbott {\it et al.} [LIGO Scientific and Virgo Collaborations],
  Phys.\ Rev.\ Lett.\  {\bf 119} (2017) no.16,  161101.

\bibitem{Ashtekar:2005qt}
  A.~Ashtekar and M.~Bojowald,
  Class.\ Quant.\ Grav.\  {\bf 23} (2006) 391.

\bibitem{Nicolini:2005vd}
  P.~Nicolini, A.~Smailagic and E.~Spallucci,
  Phys.\ Lett.\ B {\bf 632} (2006) 547.

\bibitem{LopezDominguez:2006wd}
  J.~C.~L\'opez-Dom\'inguez, O.~Obreg\'on, M.~Sabido and C.~Ramirez,
  Phys.\ Rev.\ D {\bf 74} (2006) 084024.

\bibitem{Hossenfelder:2009fc}
  S.~Hossenfelder, L.~Modesto and I.~Pr\'emont-Schwarz,
  Phys.\ Rev.\ D {\bf 81} (2010) 044036.

\bibitem{Bojowald:2018xxu}
  M.~Bojowald, S.~Brahma and D.~h.~Yeom,
  Phys.\ Rev.\ D {\bf 98} (2018) no.4,  046015.

\bibitem{BenAchour:2018khr}
  J.~Ben Achour, F.~Lamy, H.~Liu and K.~Noui,
  EPL {\bf 123} (2018) no.2,  20006.

\bibitem{Ashtekar:2018lag}
  A.~Ashtekar, J.~Olmedo and P.~Singh,
  Phys.\ Rev.\ Lett.\  {\bf 121} (2018) no.24,  241301.

\bibitem{Ashtekar:2018cay}
  A.~Ashtekar, J.~Olmedo and P.~Singh,
  Phys.\ Rev.\ D {\bf 98} (2018) no.12,  126003.

\bibitem{Bodendorfer:2019cyv}
  N.~Bodendorfer, F.~M.~Mele and J.~M\"unch,
  arXiv:1902.04542 [gr-qc].

\bibitem{Capozziello:2011et}
  S.~Capozziello and M.~De Laurentis,
  Phys.\ Rept.\  {\bf 509} (2011) 167.

\bibitem{Englert:1976ep}
  F.~Englert, C.~Truffin and R.~Gastmans,
  Nucl.\ Phys.\ B {\bf 117} (1976) 407.

\bibitem{narlikar:1977nf}
  J.~v.~narlikar and A.~k.~kembhavi,
  Lett.\ Nuovo Cim.\  {\bf 19} (1977) 517.

\bibitem{tHooft:2011aa}
  G.~'t Hooft,
  Found.\ Phys.\  {\bf 41} (2011) 1829.

\bibitem{Dabrowski:2008kx}
  M.~P.~D\c{a}browski, J.~Garecki and D.~B.~Blaschke,
  Annalen Phys.\  {\bf 18} (2009) 13.

\bibitem{Mannheim:2011ds}
  P.~D.~Mannheim,
  Found.\ Phys.\  {\bf 42} (2012) 388.

\bibitem{Mannheim:2016lnx}
  P.~D.~Mannheim,
  Prog.\ Part.\ Nucl.\ Phys.\  {\bf 94} (2017) 125.

\bibitem{Modesto:2016max}
  L.~Modesto and L.~Rachwa\l,
  arXiv:1605.04173 [hep-th].

\bibitem{Bambi:2016wdn}
C.~Bambi, L.~Modesto and L.~Rachwa\l,
  JCAP {\bf 1705} (2017) no.05,  003.

\bibitem{Bambi:2017yoz}
  C.~Bambi, Z.~Cao and L.~Modesto,
  Phys.\ Rev.\ D {\bf 95} (2017) no.6,  064006.

\bibitem{Zhou:2018bxk}
  M.~Zhou, Z.~Cao, A.~Abdikamalov, D.~Ayzenberg, C.~Bambi, L.~Modesto and S.~Nampalliwar,
  Phys.\ Rev.\ D {\bf 98} (2018) no.2,  024007.

\bibitem{Toshmatov:2017bpx}
  B.~Toshmatov, C.~Bambi, B.~Ahmedov, Z.~Stuchl\'ik and J.~Schee,
  Phys.\ Rev.\ D {\bf 96} (2017) 064028.

\bibitem{Bambi:2016yne}
  C.~Bambi, L.~Modesto, S.~Porey and L.~Rachwa\l,
  JCAP {\bf 1709} (2017) no.09,  033.

\bibitem{Bambi:2017ott}
  C.~Bambi, L.~Modesto, S.~Porey and L.~Rachwa\l,
  Eur.\ Phys.\ J.\ C {\bf 78} (2018) no.2,  116.

\bibitem{Turimov:2018pey}
  B.~Turimov, B.~Ahmedov, A.~Abdujabbarov and C.~Bambi,
  Phys.\ Rev.\ D {\bf 97} (2018) 124005.

\bibitem{Chakrabarty:2017ysw}
  H.~Chakrabarty, C.~A.~Benavides-Gallego, C.~Bambi and L.~Modesto,
  JHEP {\bf 1803} (2018) 013.

\bibitem{Toshmatov:2017kmw}
  B.~Toshmatov, C.~Bambi, B.~Ahmedov, A.~Abdujabbarov and Z.~Stuchl\'ik,
  Eur.\ Phys.\ J.\ C {\bf 77} (2017) no.8,  542.

\bibitem{Zhang:2018qdk}
  Q.~Zhang, L.~Modesto and C.~Bambi,
  Eur.\ Phys.\ J.\ C {\bf 78} (2018) no.6,  506.

\bibitem{Zhang:2017amt}
  H.~Zhang, Y.~Zhang and X.~Z.~Li,
  Nucl.\ Phys.\ B {\bf 921} (2017) 522.

\bibitem{Schutz:1985zz}
  B.~F.~Schutz and C.~M.~Will,
  Astrophys.\ J.\  {\bf 291} (1985) L33.
  
\bibitem{Iyer:1986np}
  S.~Iyer and C.~M.~Will,
  Phys.\ Rev.\ D {\bf 35} (1987) 3621.
  
\bibitem{Konoplya:2003ii}
  R.~A.~Konoplya,
  Phys.\ Rev.\ D {\bf 68} (2003) 024018.
  
\bibitem{Matyjasek:2017psv}
  J.~Matyjasek and M.~Opala,
  Phys.\ Rev.\ D {\bf 96} (2017) no.2,  024011.









\bibitem{Kobayashi:2012kh}
  T.~Kobayashi, H.~Motohashi and T.~Suyama,
  Phys.\ Rev.\ D {\bf 85} (2012) 084025
   Erratum: [Phys.\ Rev.\ D {\bf 96} (2017) no.10,  109903].

\bibitem{Kobayashi:2014wsa}
  T.~Kobayashi, H.~Motohashi and T.~Suyama,
  Phys.\ Rev.\ D {\bf 89} (2014) no.8,  084042.

\bibitem{Minamitsuji:2014hha}
  M.~Minamitsuji,
  Gen.\ Rel.\ Grav.\  {\bf 46} (2014) 1785.

\bibitem{Dong:2017toi}
  R.~Dong, J.~Sakstein and D.~Stojkovic,
  Phys.\ Rev.\ D {\bf 96} (2017) no.6,  064048.
  
\bibitem{Tattersall:2018nve}
  O.~J.~Tattersall and P.~G.~Ferreira,
  Phys.\ Rev.\ D {\bf 97} (2018) no.10,  104047.

\bibitem{Sebastian:2014qra}
  S.~Sebastian and V.~C.~Kuriakose,
  Mod.\ Phys.\ Lett.\ A {\bf 30} (2015) no.05,  1550012.

\bibitem{Bhattacharyya:2017tyc}
  S.~Bhattacharyya and S.~Shankaranarayanan,
  Phys.\ Rev.\ D {\bf 96} (2017) no.6,  064044.

\bibitem{Bhattacharyya:2018qbe}
  S.~Bhattacharyya and S.~Shankaranarayanan,
  Eur.\ Phys.\ J.\ C {\bf 78} (2018) no.9,  737.

\bibitem{Chen:2018mkf}
  C.~Y.~Chen and P.~Chen,
  Phys.\ Rev.\ D {\bf 98} (2018) no.4,  044042.

\bibitem{Chen:2018vuw}
  C.~Y.~Chen, M.~Bouhmadi-L\'opez and P.~Chen,
  Eur.\ Phys.\ J.\ C {\bf 79} (2019) 63.

\bibitem{Fernando:2014gda}
  S.~Fernando and T.~Clark,
  Gen.\ Rel.\ Grav.\  {\bf 46} (2014) no.12,  1834.

\bibitem{Prasia:2016fcc}
  P.~Prasia and V.~C.~Kuriakose,
  Gen.\ Rel.\ Grav.\  {\bf 48} (2016) no.7,  89.
  
\bibitem{Blazquez-Salcedo:2016enn}
  J.~L.~Bl\'azquez-Salcedo, C.~F.~B.~Macedo, V.~Cardoso, V.~Ferrari, L.~Gualtieri, F.~S.~Khoo, J.~Kunz and P.~Pani,
  Phys.\ Rev.\ D {\bf 94} (2016) no.10,  104024.

\bibitem{Blazquez-Salcedo:2016yka}
  J.~L.~Bl\'azquez-Salcedo {\it et al.},
  IAU Symp.\  {\bf 324} (2016) 265.

\bibitem{Blazquez-Salcedo:2017txk}
  J.~L.~Bl\'azquez-Salcedo, F.~S.~Khoo and J.~Kunz,
  Phys.\ Rev.\ D {\bf 96} (2017) no.6,  064008.

\bibitem{Blazquez-Salcedo:2018pxo}
  J.~L.~Bl\'azquez-Salcedo, Z.~A.~Motahar, D.~D.~Doneva, F.~S.~Khoo, J.~Kunz, S.~Mojica, K.~V.~Staykov and S.~S.~Yazadjiev,
  arXiv:1810.09432 [gr-qc].

\bibitem{Toshmatov:2016bsb}
  B.~Toshmatov, Z.~Stuchl\'ik, J.~Schee and B.~Ahmedov,
  Phys.\ Rev.\ D {\bf 93} (2016) no.12,  124017.

\bibitem{Lin:2016wci}
  K.~Lin, W.~L.~Qian and A.~B.~Pavan,
  Phys.\ Rev.\ D {\bf 94} (2016) no.6,  064050.

\bibitem{Cardoso:2003qd}
  V.~Cardoso, J.~P.~S.~Lemos and S.~Yoshida,
  JHEP {\bf 0312} (2003) 041.

\bibitem{Cardoso:2003vt}
  V.~Cardoso, J.~P.~S.~Lemos and S.~Yoshida,
  Phys.\ Rev.\ D {\bf 69} (2004) 044004.

\bibitem{Cardoso:2004cj}
  V.~Cardoso, G.~Siopsis and S.~Yoshida,
  Phys.\ Rev.\ D {\bf 71} (2005) 024019.

\bibitem{Ding:2017gfw}
  C.~Ding,
  Phys.\ Rev.\ D {\bf 96} (2017) no.10,  104021.

\bibitem{Nollert:1999ji}
  H.~P.~Nollert,
  Class.\ Quant.\ Grav.\  {\bf 16} (1999) R159.

\bibitem{Berti:2009kk}
  E.~Berti, V.~Cardoso and A.~O.~Starinets,
  Class.\ Quant.\ Grav.\  {\bf 26} (2009) 163001.

\bibitem{Konoplya:2011qq}
  R.~A.~Konoplya and A.~Zhidenko,
  Rev.\ Mod.\ Phys.\  {\bf 83} (2011) 793.

\bibitem{Berti:2015itd}
  E.~Berti {\it et al.},
  Class.\ Quant.\ Grav.\  {\bf 32} (2015) 243001.

\bibitem{Cardoso:2019mqo}
  V.~Cardoso, M.~Kimura, A.~Maselli, E.~Berti, C.~F.~B.~Macedo and R.~McManus,
  arXiv:1901.01265 [gr-qc].

\bibitem{Chandrabook}
S.~Chandrasekhar(ed.): The Mathematical Theory of Black Holes. Oxford University Press,
Oxford (1992).

\bibitem{Regge:1957td}
  T.~Regge and J.~A.~Wheeler,
  Phys.\ Rev.\  {\bf 108} (1957) 1063.

\bibitem{Leaver:1986gd}
  E.~W.~Leaver,
  Phys.\ Rev.\ D {\bf 34} (1986) 384.

\bibitem{Jansen:2017oag}
  A.~Jansen,
  Eur.\ Phys.\ J.\ Plus {\bf 132} (2017) no.12,  546.

\bibitem{Cardoso:2008bp}
  V.~Cardoso, A.~S.~Miranda, E.~Berti, H.~Witek and V.~T.~Zanchin,
  Phys.\ Rev.\ D {\bf 79} (2009) 064016.
  
\bibitem{Konoplya:2017wot}
  R.~A.~Konoplya and Z.~Stuchl\'ik,
  Phys.\ Lett.\ B {\bf 771} (2017) 597.
  
\bibitem{Toshmatov:2018tyo}
  B.~Toshmatov, Z.~Stuchl\'ik, J.~Schee and B.~Ahmedov,
  Phys.\ Rev.\ D {\bf 97} (2018) no.8,  084058.

\bibitem{Gundlach:1993tp}
  C.~Gundlach, R.~H.~Price and J.~Pullin,
  Phys.\ Rev.\ D {\bf 49} (1994) 883.

\bibitem{Chirenti:2007mk}
  C.~B.~M.~H.~Chirenti and L.~Rezzolla,
  Class.\ Quant.\ Grav.\  {\bf 24} (2007) 4191.

\bibitem{Aneesh:2018hlp}
  S.~Aneesh, S.~Bose and S.~Kar,
  Phys.\ Rev.\ D {\bf 97} (2018) no.12,  124004.


 \bibitem{Abramow}
  M.~Abramowitz and I.~Stegun, \textit{Handbook on Mathematical Functions} (Dover, 1980).
  
\end{thebibliography}
\end{document}